\documentclass{aa} 

\usepackage{graphicx}
\usepackage{listings}
\usepackage{makecell}
\usepackage{txfonts}
\usepackage{bm}
\newcommand{\idefix}{\textsc{Idefix}\xspace}
\newcommand{\pluto}{\textsc{Pluto}\xspace}
\newcommand{\kokkos}{\textsc{Kokkos}\xspace}
\newcommand{\Ramses}{\textsc{Ramses}\xspace}
\newcommand{\Athena}{\textsc{Athena}\xspace}
\newcommand{\Dedalus}{\textsc{Dedalus}\xspace}
\newcommand{\Snoopy}{\textsc{Snoopy}\xspace}
\newcommand{\Magic}{\textsc{Magic}\xspace}
\newcommand{\Nirvana}{\textsc{Nirvana}\xspace}
\newcommand{\Zeus}{\textsc{Zeus}\xspace}
\newcommand{\Pencil}{\textsc{Pencil}\xspace}
\newcommand{\Fargo}{\textsc{Fargo}\xspace}
\newcommand{\FargoTD}{\textsc{Fargo3d}\xspace}
\newcommand{\emf}{\bm{\mathcal{E}}}
\newcommand{\tensor}[1]{\overline{\overline{#1}}}
\newcommand{\etaO}{\eta_\mathrm{O}}
\newcommand{\xH}{x_\mathrm{H}}
\newcommand{\xAD}{x_\mathrm{AD}}

\begin{document}

   \title{IDEFIX: a versatile performance-portable Godunov code for astrophysical flows}

   \author{G. R. J. Lesur
          \and
          S. Baghdadi
          \and
          G. Wafflard-Fernandez
          \and
          J. Mauxion
          \and
          C. M. T. Robert
          \and 
          M. Van den Bossche
          }

   \institute{Univ. Grenoble Alpes, CNRS, IPAG, 38000 Grenoble, France\\
              \email{geoffroy.lesur@univ-grenoble-alpes.fr}
                      }

   \date{Received September 15, 1996; accepted March 16, 1997}

 
  \abstract
   {Exascale super-computers now becoming available rely on hybrid energy-efficient architectures that involve an accelerator such as Graphics Processing Units (GPU). Leveraging the computational power of these machines often means a significant rewrite of the numerical tools each time a new architecture becomes available.}
   {We present \idefix, a new code for astrophysical flows that relies on the \kokkos meta-programming library to guarantee performance portability on a wide variety of architectures while keeping the code as simple as possible to the user.}
   {\idefix is based on a Godunov finite-volume method that solves the non-relativistic HD and MHD equations on various grid geometries. \idefix includes a large choice of solvers and several additional modules (constrained transport, orbital advection, non-ideal MHD) allowing users to address complex astrophysical problems. }
   {\idefix has been successfully tested on Intel and AMD CPUs (up to 131\,072 CPU cores on Irene-Rome at TGCC) as well as NVidia and AMD GPUs (up to 1024 GPUs on Adastra at CINES). \idefix achieves more than $10^8$ cell/s in MHD on a single NVidia V100 GPU and $3\times 10^{11}$ cell/s on 256 Adastra nodes (1024 GPUs) with 95\% parallelisation efficiency (compared to single node). For the same problem, \idefix is up to 6 times more energy efficient on GPUs compared to Intel Cascade Lake CPUs.}
   {\idefix is now a mature exascale-ready open-source code that can be used on a large variety of astrophysical and fluid dynamics applications.}


   \maketitle
%

\section{Introduction}
The development of modern theoretical astrophysics relies heavily on numerical modelling because of the complex interplay between several and often non-linear physical processes. This is especially true for many astrophysical flows ranging from stellar interiors to large-scale galaxy clusters which all rely on multi-scale modelling of the plasma. Many techniques have been designed to model astrophysical flows, the most popular being based on a fixed grid approach: finite differences schemes, with codes like \Zeus \citep{Stone.Norman92a} and its derivatives like \Fargo \citep{Masset00} or the \Pencil code \citep{PencilCodeCollaboration.Brandenburg.ea21}, finite volume and in particular Godunov-like shock-capturing schemes like \Ramses \citep{Teyssier02, Fromang.Hennebelle.ea06} \Athena \citep{Stone.Gardiner.ea08,Stone.Tomida.ea20}, \pluto \citep{Mignone.Bodo.ea07}, \Nirvana \citep{Ziegler08} and spectral methods often used in quasi-incompressible problems like \Magic \citep{Wicht02a,Gastine.Wicht12}, \Snoopy \citep{Lesur.Longaretti07} or \Dedalus \citep{Burns.Vasil.ea20}. All of these codes are largely used in the community and are usually written in Fortran or C for ease of use. 

With a few exceptions, these codes are designed to run on CPU-based systems like x86 machines, with typically a few tens of independent cores in each CPUs that are inter-connected with a high-bandwidth and low-latency network. The advent of exascale supercomputers and the requirement of energy sobriety in high-performance computing is now driving a shift towards hybrid machines. Clusters of CPUs are progressively replaced by energy-efficient architectures, often based on manycore systems such as GPUs (Graphics Processing Units). In this approach, each GPU is made of thousands of cores which are all working synchronously following the SIMD (Single Instruction Multiple Data) paradigm. This change in hardware architecture means that codes should at the very least be ported to these new architectures. To make things worse, each hardware manufacturer tends to push its own programming language or library for its own hardware, often as an extension of Fortran or C (NVidia Cuda, AMD HIP Rocm or Apple Metal to name a few).

This has a dramatic impact on the numerical astrophysics community. It is already a long process to write a new code, but support and maintenance are often scarce because of the limited staff available, and the low recognition of support activities.  It is therefore almost impossible to develop and maintain $n$ versions of a code for $n$ architecture variants given the human resources available. Nevertheless, our community must find ways to exploit these architectures as they will be dominant in the future, not only in exascale machines but also in university clusters because of constraints on energy sobriety.

To address some of these issues, we have developed a new code based on a high-order Godunov scheme, that relies on the \kokkos \citep{Trott.Lebrun-Grandie.ea22} metaprogramming library. This allows us to exploit a performance-portability approach, where a single source code can be compiled for a large variety of target architectures, with minimal impact on performance. While our original intention was to port the \pluto code, we reckoned that it was more efficient in the long run to write a code from scratch using modern C++. This allowed us to leverage class encapsulation, polymorphism, exception handling and C++ containers that are all absent from the C standard. We present the main features of this code in this paper. We first introduce the basics of the algorithm with the various physical modules we have implemented in section 2. We then focus on implementation details. The various standard tests used to validate the code are presented in section 4. Finally, we discuss the code performances for various target architectures in section 5, as well as future prospects for this tool.

\section{\label{sec:algo}Algorithm}
\subsection{Equations}
In all generality, the dimensionless equations solved by \idefix read:
\begin{align}
\label{eq:cont}\partial_t\rho+\bm{\nabla\cdot}\big[\rho\bm{v}\big]&=0,\\
\label{eq:mom}\partial_t(\rho\bm{v})+\bm{\nabla \cdot}\big[\rho \bm{v}\otimes \bm{v}-\bm{B}\otimes\bm{B}+\mathcal{P}\mathbb{I}+\tensor{\Pi}\big]&=-\rho \bm{\nabla}\psi,\qquad\\
\nonumber \partial_t \bm{B}+\bm{\nabla\times}\bigg[-\bm{v\times B}+\etaO\bm{J}+\xH\bm{J\times B}\qquad &\\
\label{eq:ind}-\xAD(\bm{J\times B})\bm{\times B}\Big] &=0,\\
\label{eq:energy}\nonumber\partial_tE+\bm{\nabla \cdot}\bigg[(E+\mathcal{P})\bm{v}-\bm{B}(\bm{B\cdot v})+\tensor{\Pi}\bm{\cdot v}&\\
\nonumber+ \etaO \bm{J\times B}+ \xH(\bm{J\times B})\bm{\times B}\quad&\\
 \qquad- \xAD\big[(\bm{J\times B})\bm{\times B}\big]\bm{\times B}
+\kappa\bm{\nabla} T\bigg]&=-\rho\bm{v\cdot\nabla}\psi.
\end{align}
Equations (\ref{eq:cont}) and (\ref{eq:mom}) are the continuity and momentum equations, where $\rho$ is the flow density, $\bm{v}$ is the velocity, $\bm{B}$ is the magnetic field, $\mathcal{P}\equiv P+B^2/2$ is the generalised pressure with $P$ the gas pressure, $\psi$ is the gravitational potential, $\mathbb{I}$ is the identity tensor and $\tensor{\Pi}$ is the viscous stress tensor, defined as
\begin{align*}
\tensor{\Pi}\equiv \eta_1\bigg(\bm{\nabla v}+(\bm{\nabla v})^{T}\bigg)+\bigg(\eta_2-\frac{2}{3}\eta_1\bigg)\bm{\nabla\cdot v}\mathbb{I}	
\end{align*}
where $\eta_1$ and $\eta_2$ are the standard and bulk viscosities. In magneto-hydrodynamics (MHD), this is complemented by the induction equation (\ref{eq:ind}), where $\etaO$ is the Ohmic diffusivity, $\bm{J}\equiv\bm{\nabla\times B}$ is the electrical current, $\xH$ is the Hall parameter and $\xAD$ is the ambipolar diffusion parameter. This set of equations is completed by the total energy equation (\ref{eq:energy}) where $E$ is the total energy density
\begin{align*}
E=e+\frac{\rho v^2}{2}+\frac{B^2}{2}	,
\end{align*}
$e$ being the internal energy density, and $\kappa$ the thermal diffusion coefficient (assumed to be a scalar).

Note that \idefix does not assume any explicit physical units. The user is free to choose its physical units depending on the problem at hand, keeping in mind that the equations effectively solved by the code are the ones shown above.
 
\subsection{Method of solution}
\begin{figure*}
   \centering
   \includegraphics[width=\hsize]{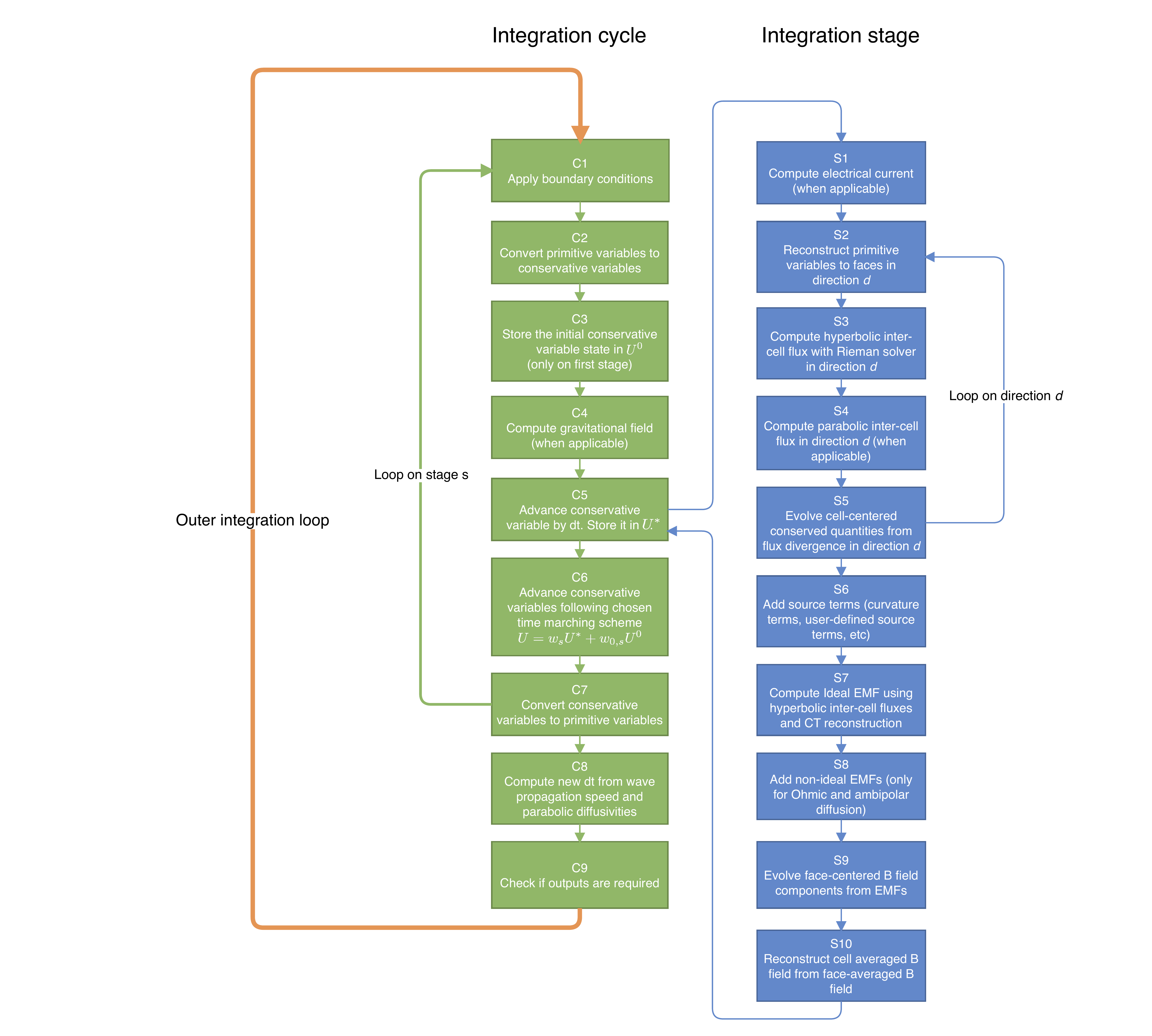}
      \caption{Schematics of the integration algorithm in \idefix}
         \label{fig:idefix-algo}
   \end{figure*}

This set of conservative equations can be reformulated using a generalised vector of conserved quantities $U=(\rho,\rho\bm{v},\bm{B},E)$ as
\begin{align}
\label{eq:gal}
	\partial_t U + \bm{\nabla\cdot}\bm{F} = S
\end{align}
where $\bm F$ is the flux of conserved quantity $U$, and $S$ is a source term which cannot be included in the divergence term. \idefix is fundamentally a Godunov-type finite-volume code, designed to model astrophysical plasmas. Hence, \idefix primarily evolves the cell-averaged conserved quantities $\langle U\rangle$ which are computed using the face- and time-averaged fluxes $\langle F \rangle$ \citep{Toro09}. While we omit the $\langle\quad\rangle$ symbol in the following of the manuscript for simplicity, this distinction should be kept in mind. The algorithm is fairly standard and follows the reconstruct-solve-average (RSA) approach similarly to other codes available on the market such as \pluto, \Athena, \Ramses and \Nirvana, to name a few. The implementation design of this algorithm in \idefix is intentionally close to that of the \pluto code to simplify the portability of physical setups from one code to the other (see section \ref{sec:Pluto}). Hence, we will not go into the details of the algorithm which may be found in the papers mentioned above but instead focus on the main stages of the algorithms, and the differences with \pluto. The full algorithm with its various steps and loops is represented graphically in fig.~\ref{fig:idefix-algo} for reference.

\subsection{\label{sec:hyp}Hyperbolic hydrodynamical terms}
As with all Godunov codes, $\idefix$ uses a strategy based on cell reconstruction followed by inter-cell flux computation using Riemann solvers to follow the evolution of $U$ due to hyperbolic terms. In $\idefix$, the reconstruction to cell faces (step S2) is performed on the primitive variables $(\rho, \bm{v}, \bm{B}, P)$. In doing so, we avoid the decomposition in characteristics used in other codes while keeping the basic physical constraints (density and pressure positivity for instance). Reconstruction can be performed using either a flat reconstruction (1st order accurate), piecewise linear reconstruction using the van Leer slope limiter (2nd order accurate), compact 3rd order reconstruction (up to 4th order accurate \citealt{Cada.Torrilhon09}) and piecewise parabolic reconstruction using an extremum-preserving limiter (4th order accurate on regularly spaced grids, \citealt{Colella.Woodward84,Colella.Sekora08})\footnote{Note however that the code is still at most globally second order, see \S\ref{sec:mhd-tests}}. The number of ghost cells is automatically adjusted to the requirements of the reconstruction scheme.

Once the reconstruction on cell faces is done, a Riemann solver is used to compute the inter-cell flux (step S3). \idefix can use, ordered from most to least diffusive, the Lax-Friedrichs Russanov flux \citep{Rusanov62}, the Harten, Lax, and van Leer (HLL) approximate solver \citep{Einfeldt.Roe.ea91}, the HLLC solver (a version of the HLL solver that keeps the contact wave, \citealt{Toro.Spruce.ea94}) and the approximate linear Roe Riemann solver \citep{Cargo.Gallice97}. In the MHD case, the HLLC solver is replaced by its magnetised HLLD counterpart \citep{Miyoshi.Kusano05}. These Riemann solvers can be chosen freely at run time, allowing the user to experiment with each of them. Note that, in contrast to \pluto which treats pressure gradients as a source term in the momentum equation, pressure in \idefix is included in the momentum flux, and hence solved for by the Riemann solver. This approach has the advantage of treating all of the variables equally but creates additional pressure source terms in non-cartesian geometries, which are absent from \pluto. As a consequence, the results of \pluto and \idefix are expected to differ, even if the same algorithm combination is chosen.

\subsection{\label{sec:mhd}Magneto-hydrodynamics}

When the MHD module is switched on, \idefix adds the induction equation (\ref{eq:ind}) to the set of equations being solved, which can be written as:
\begin{align}
\label{eq:ind_short}
    \partial_t \bm{B}=-\bm{\nabla\times \mathcal{E}}
\end{align}
where $\bm{\mathcal{E}}$ is the induced electric field.
In contrast to the discretisation of (\ref{eq:gal}) that uses cell-averaged quantities, the induction equation uses face-averaged magnetic field components $B$ and evolves them using the Constrained Transport scheme (CT, \citealt{Evans.Hawley88}). In this approach, the electric field $\bm{\mathcal{E}}$ is discretised on cell edges, and a discrete version of Stokes theorem is used to evolve $\bm{B}$ from the contour integral of $\mathcal{E}$. This approach in principle allows us to keep $\bm{\nabla\cdot B}=0$ at machine precision. Note however that even with CT, roundoff errors can accumulate in $B$, leading to potentially significant $\nabla\cdot \bm{B}$. This can be encountered for particularly long simulations running for billions of timesteps, and even more so when a Runge-Kutta-Legendre super-time-stepping scheme is used in conjunction with CT (see \S\ref{sec:rkl}). Hence, $\idefix$ allows the user to split the induction equation into two parts
\begin{align}
\label{eq:ind_vec}
    \partial_t \bm{A}&=-\bm{\mathcal{E}},\\
    \bm{B}&=\bm{\nabla\times A}.
\end{align}
By defining the vector potential $\bm{A}$ on cell edges, the resulting algorithm is mathematically equivalent to the original CT algorithm. However, it prevents roundoff errors from accumulating in $\bm{\nabla\cdot B}$, guaranteeing that the solenoidal condition is \emph{always} satisfied, even at late times. The only drawback of this vector potential approach is the extra cost to store $\bm{A}$ in memory, and the need to code an initial condition for $\bm{A}$ instead of $\bm{B}$. Since our gauge choice is so that $\bm{A}=\int\bm{\mathcal{E}}\,\mathrm{d}t$, there is a possibility of overflowing the vector potential in physical problems where a large-scale constant EMF is present. While we have not encountered this difficulty in real-life application, it is recommended to clean the vector potential periodically using an adequate gauge transformation if such a situation is expected to occur. Note finally that this vector potential formulation is always available, even when orbital advection, non-ideal MHD and/or super time-stepping is used. 

The main difficulty with the CT scheme is to reconstruct $\emf$ on cell edges in a way that is numerically stable and mathematically consistent with steps S2---S4. Two classes of schemes in \idefix satisfy these requirements and are performed in S7: schemes that rely on the face-centred fluxes computed by the Riemann solvers in S3, following \cite{Gardiner.Stone05}, and schemes that solve an additional 2D Riemann problem at each cell edge as proposed by \cite{Londrillo.delZanna04}. In the first category, \idefix can use the arithmetic average of fluxes, and the $\mathcal{E}^0$ and $\mathcal{E}^c$ schemes of \citealt{Gardiner.Stone05}. In the second category, \idefix can use the 2D HLL and HLLD 2D Riemann solvers following the method of \cite{Mignone.DelZanna21}. By default, \idefix uses the $\mathcal{E}^c$ scheme, similar to \Athena, which is a good trade-off between accuracy, stability and computing time. The choice of the $\emf$ reconstruction scheme can be changed at runtime.

\subsection{\label{sec:par}Parabolic terms}

Several parabolic terms have been implemented in \idefix: viscous diffusion, thermal diffusion, Ohmic diffusion, ambipolar diffusion and the Hall effect. These terms are added to the hyperbolic fluxes in S4 so as not to break the numerical conservation of $U$. For MHD diffusive parabolic terms (Ohmic and ambipolar diffusion), we add their contribution using a centred finite-difference formula to $\mathcal{E}$ before advancing $\bm{B}$ (or $\bm{A}$) in S8. Finally, the Hall effect requires special treatment. This term is notoriously difficult to implement owing to its dispersive but non-dissipative nature. In \idefix, we follow the approach presented in \cite{Lesur.Kunz.ea14} and add the whistler speed to the Riemann fan of the HLL solver in S3. We further improve this approach by using the whistler wave speed only to evolve the magnetic field fluxes, and keep the usual ideal MHD wave speeds for the other components of $F$, following \cite{Marchand.Tomida.ea19}. This treatment reduces the numerical diffusion on non-magnetic conserved quantities. Finally, $\emf$ is reconstructed on cell edges using an arithmetic average of $F$, as in \cite{Lesur.Kunz.ea14}. We tried modifying the $\mathcal{E}^c$ reconstruction scheme to include the Hall term but found that this approach led to numerical instabilities. While known to be diffusive, our implementation of the Hall effect guarantees the stability of the resulting scheme.

\subsection{Grid and Geometry}

\idefix can run problems in cartesian, polar/cylindrical and spherical geometry. In all of these geometries, the grid is rectilinear, but its spacing may be parameterized as piecewise functions of coordinates. While this approach is clearly not as flexible as adaptive mesh refinement, it allows refinement  around regions of interest, at a reduced computational cost and code complexity. In polar and spherical geometry, we choose the conserved quantity in the azimuthal direction to be the vertical component of the angular momentum. This guarantees conservation of angular momentum to machine precision. Great care has been taken for spherical domains extending up to the axis. A special "axis" boundary condition is provided which guarantees regularity of solutions around the axis, following \cite{Zhu.Stone18}. 

\subsection{\label{sec:timeIntegrator}Time integrator}

The outermost time integration loop is called a cycle. At each cycle $n$, the solution $U^{(n)}$ is advanced by a time step $dt$ giving $U^{(n+1)}$. Within a cycle, several time-marching schemes can be invoked to evolve physical quantities, each of them possibly using several sub-cycles called stages. In \idefix, the default time-integrator is an explicit multi-stage scheme. To guarantee a given level of accuracy, \idefix supports several orders for this scheme, following the method of lines: Euler (1st order in time) and total variation diminishing Runge-Kutta 2 and 3 (respectively 2nd and 3rd order, \citealt{Gottlieb.Shu98}). 

Each explicit time-integration cycle starts with the solution at time $t$, $U_0\equiv U^{(n)}$. For each stage $s$, \idefix advances the previous stage $U_{s-1}$ by $\mathrm{d}t$ using the algorithm described in \S\ref{sec:hyp}---\S\ref{sec:par} (steps S1---S10, which we note in a single operator $\mathcal{S}_{\mathrm{d}t}$) and store it in $U^*=\mathcal{S}_{\mathrm{d}t}(U_{s-1})$. This solution is then linearly combined with $U_0$ to give $U_s$. Formally, for each stage $s\ge 1$, we compute
\begin{align}
U_s = w_sU^* +w^0_sU_0,
\end{align}
where $w_s$ and $w^0_s$ are fixed real coefficients (Tab.~\ref{tab:weights}). 
\begin{table}
\begin{center}
\caption{\label{tab:weights}Weight coefficients for the time-integration schemes available in \idefix. Note that the Euler scheme is a single-stage scheme, while RK2 and RK3 are respectively two and third stages schemes}
\begin{tabular}{|c|r|r|r|}
\hline
Weight & 	    Euler & RK2 & RK3 \\
\hline
$w_1$ &      1    &             1 & 1             \\
$w^0_1$ &    0    &            0 & 0             \\
\hline
$w_2$&          &     1/2      & 1/4           \\
$w^0_2$&        &     1/2      & 3/4           \\
\hline
$w_3$&         &            & 2/3\\
$w^0_3$&        &            & 1/3\\
\hline
\end{tabular}
\end{center}

\end{table}

Since the main time marching scheme in \idefix is explicit, a CFL \citep{Courant.Friedrichs.ea28} condition is needed to guarantee numerical stability. This constraint is computed as a generalisation of \cite{Beckers92}, eq. 62:
\begin{align}
\label{eq:cfl}
    dt= \sigma_\mathrm{CFL}\left(\max_{\mathcal{V}}\left[\sum_d\frac{c_{\mathrm{max},d}}{d\ell_d}+\frac{2\eta_\mathrm{max}}{d\ell_d^2}\right]\right)^{-1}
\end{align}
where $\sigma_\mathrm{CFL}<1$ is a safety factor, $c_{\mathrm{max},d}$ is the maximum signal speed in current cell in direction $d$ (computed in the Riemann solver step S3), $\eta_\mathrm{max}$ is the maximum diffusion coefficient in current cell and $d\ell_d$ is the current cell length in direction $d$. Note that the maximum in the above formula is taken over the entire integration volume $\mathcal{V}$, while the summation is performed in all directions of integration $d$. Finally, we emphasize that our CFL condition is different from that of \pluto \citep[][eq. 7]{Mignone.Flock.ea12} in at least two ways: 1- we do not include any pre-factor on the number of dimensions, so that $\sigma_\mathrm{CFL}= D \times C_{a,\pluto}$ where $D$ is the number of dimensions so that the stability condition is always $\sigma_\mathrm{CFL}<1$ in \idefix and 2- our CFL condition combines the point-wise hyperbolic and parabolic constraints, while \pluto computes a global $dt$ for hyperbolic terms and parabolic terms separately. The end result is that $dt$ computed by \idefix can be somewhat larger than the more restrictive time step from \pluto, while still satisfying \cite{Beckers92}'s constraint on a point-wise basis.  

\subsection{Super time-stepping: the Runge-Kutta-Legendre scheme\label{sec:rkl}}
In some circumstances, for instance, in very diffusive plasmas, the CFL condition (\ref{eq:cfl}) on parabolic terms leads to prohibitively low $dt$. In such cases, it is recommended to use super time-stepping to integrate specifically this term and remove it from the standard explicit scheme. In \idefix, super time-stepping is implemented using the explicit Runge-Kutta-Legendre scheme (RKL) from \cite{Meyer.Balsara.ea14} and \cite{Vaidya.Prasad.ea17}. The advantage of the RKL scheme compared to the usual super time-stepping based on Chebyshev polynomials (sometimes called RKC) is its increased accuracy and robustness, and the absence of any free ``damping'' parameter as the method is intrinsically stable. In \idefix, we have implemented a second-order RKL scheme (a first-order scheme is also available for debugging purposes). Splitting the time-step (\ref{eq:cfl}) into a hyperbolic and parabolic part
\begin{align}
\label{eq:cfl_sts}
    dt_\mathrm{hyp}= \left(\max_{\mathcal{V}}\left[\sum_d\frac{c_{\mathrm{max},d}}{d\ell_d}\right]\right)^{-1},\\
    dt_\mathrm{par}= \left(\max_{\mathcal{V}}\left[\sum_d\frac{2\eta_\mathrm{max}}{d\ell_d^2}\right]\right)^{-1},
\end{align}
then the RKL scheme saves computational time when $dt_\mathrm{hyp}\gg dt_\mathrm{par}$. More precisely, the number of stages $s_\mathrm{RKL}$ required by the second order RKL scheme satisfies \citep{Meyer.Balsara.ea14}:
\begin{align}
dt_\mathrm{hyp}=dt_\mathrm{par}\frac{(s_\mathrm{RKL}^2+s_\mathrm{RKL}-2)}{4}.
\end{align}
Hence, when $dt_\mathrm{hyp}\gg dt_\mathrm{par}$, $s_\mathrm{RKL}\simeq 2\sqrt{dt_\mathrm{hyp}/dt_\mathrm{par}}$, which should be compared to $2dt_\mathrm{hyp}/dt_\mathrm{par}$ if the term had been integrated using the explicit RK2 scheme. 

\idefix{} allows one to use the RKL scheme for viscosity, thermal diffusion, ambipolar diffusion, Ohmic diffusion, or a combination of these terms. When enabled, the RKL scheme directly calls the same routines which are used to compute the parabolic terms in the main time integrator, so that no code is duplicated. Finally, in order to retain global second accuracy in time, the RKL stages are called alternatively before and after the main time integrator at each integration cycle.

While testing the RKL scheme for Ohmic and ambipolar diffusions, we have noted that the roundoff error on $\bm{\nabla\cdot B}$ tends to accumulate rapidly: while the standard explicit time integrator yields $\bm{\nabla\cdot B}= \mathcal{O}(\sqrt{n})$, with $n$ the number of integration cycles, we find $\bm{\nabla\cdot B}= \mathcal{O}(n)$ with the RKL scheme. This issue can lead to relatively large values of $\bm{\nabla\cdot B}$ for long simulations which may become problematic. Hence, it is recommended to use the vector potential formulation of the induction equation (\ref{eq:ind_vec}) when using RKL on a parabolic MHD term to circumvent this problem. 

\subsection{Orbital advection (Fargo-type scheme)}
When working on cold Keplerian discs (that is, when orbital velocity $\bm{U}$ is much larger than the other signal speeds involved in the system), orbital velocity is the limiting term in the CFL condition (\ref{eq:cfl}). A well-known procedure to speed up the time-integration in this case, initially proposed by \cite{Masset00} for the \textsc{Fargo} code, is to split the flux associated to the advection operator for every conserved quantity $Q$ as
\begin{align*}
	\bm{\nabla\cdot v}Q = \bm{U\cdot \nabla} Q+\bm{\nabla\cdot }\delta\bm{v}Q
\end{align*}
where we have assumed $\bm{\nabla\cdot U}=0$ and used $\delta\bm{v}\equiv \bm{v}-\bm{U}$. When $\bm{U}$ is known, the time evolution of $\partial_t Q+\bm{U\cdot \nabla} Q=0$ is analytical, so no  explicit time integration step is needed, nor any CFL condition. Hence, one splits the time integration of the full system into two stages: a numerical integration using the chosen time-marching scheme (\S\ref{sec:timeIntegrator}) using $\bm{\nabla\cdot }\delta\bm{v}Q$ in the flux function, and an analytical advection stage using the prescribed orbital velocity $\bm{U}$. The end result of this procedure is that the total velocity $v$ in the CFL condition is replaced by the residual velocity $\delta\bm{v}$, allowing for much larger time steps when $\delta v \ll v$.

In \idefix, we have implemented the conservative orbital advection scheme of \cite{Mignone.Flock.ea12}. This algorithm presents the advantage of retaining the conservation of mass, angular momentum and magnetic flux by carefully formulating the source terms implied by the operator splitting. Our implementation is fully parallelised in 3D, and in particular, allows for domain decomposition along the toroidal direction.

\section{Implementation}

\subsection{Performance portability using the \kokkos library}

\idefix has been designed with performance portability in mind, which means that a single source code can be compiled on a wide variety of target architectures and get excellent performance on each of them. To this end, \idefix uses the \kokkos library \citep{Trott.Lebrun-Grandie.ea22}. \kokkos sources are bundled as a git submodule by \idefix and are compiled alongside it. Hence, \idefix can be compiled and run out of the box.

Technically speaking, \idefix always assumes it is running on a machine that consists of a host and a device. The host is the CPU on which the code is launched, while the device can be any valid \kokkos target: CPU, GPU, \dots In special cases where no supported target is detected, the device code will be compiled as a host code, allowing \idefix to run seamlessly on systems without accelerators with very little overhead. For code portability constraints, it is always assumed that host and device memories are separated: memory owned by the host cannot be directly accessed from the device code, and vice versa. Hence, special constructs are needed to implement the algorithm described above.

\idefix provides two base constructs: arrays and parallel loops. Following the approach used in \textsc{K-Athena} \citep{Grete.Glines.ea19} arrays are stored in \verb!IdefixArrayND!\footnote{these arrays are technically aliases to \texttt{Kokkos::View}} where $1<N<4$ is the number of dimensions of the array. Parallel loops are executed using \verb!idefix_for! function, which acts similarly to \verb!for! loops in C, and seamlessly encapsulates the SIMD execution paradigm on GPUs. Hence, a usual C loop written as
\begin{lstlisting}[language=c++]
  const int N = 10;
  double array[N];
  // initialise array to 1
  for(int i = 0 ; i < N ; i++) {
    array[i] = 1.0;
  }
\end{lstlisting}
is written in \idefix as:
\begin{lstlisting}[language=c++]
  const int N = 10;
  IdefixArray1D<double> array("myArray", N);
  // initialise array to 1
  idefix_for("initLoop", 0, N, 
    KOKKOS_LAMBDA(int i) {
      array(i) = 1.0;
    }
  );
\end{lstlisting}
Note that the additional tag provided as the first argument to \verb!IdefixArray1D! and \verb!idefix_for! is only used for debugging and profiling purposes.

By construction, an \verb!IdefixArrayND! is allocated in device space and all instructions within an \verb!idefix_for! call are executed on the device. Hence, any access to elements of an  \verb!IdefixArrayND! needs to be done inside an \verb!idefix_for!. \idefix also provides arrays allocated on the host (\verb!IdefixHostArrayND!) which can be used in usual for loops, and are meant to mirror \verb!IdefixArrayND! instances. Host arrays are useful for I/O routines and complex initial conditions.

\idefix is entirely written using these constructs, hence the algorithm described in \S\ref{sec:algo} is running exclusively on the device, the host being used only for I/O, initial conditions, and handling exceptions (either triggered internally or by the user). This guarantees optimal acceleration for all combinations of algorithms provided.

\subsection{Setup portability from \pluto\label{sec:Pluto}}
\idefix's interface is very close to \pluto's. Parameter files follow the same format and use identical keywords when possible. Variable names (grid variables, hydro variables), and several preprocessor macros have been kept to simplify the portability of \pluto setups to \idefix as our research group had been using and developing physics setups for \pluto for more than a decade. Similarly, outputs (VTK\footnote{Visualization Tool Kit \citep{VTK2006}} files) follow the same format and variable naming convention as \pluto. Note however that restart dumps are not compatible between the two codes. In addition,  while \pluto relies on C arrays, \idefix is based on C++ objects and uses C++ containers. This allows for more flexibility and simpler development of physics modules.

\subsection{Inputs and outputs}

\idefix uses a text input file following \pluto's input file format that is read at runtime. Most of the code options (time integrator, Riemann solvers, physical modules) are enabled in this input file.

\idefix outputs come in two formats: an internal dump file format that contains the raw data required to restart a simulation and the VTK file format which is used for visualization and analysis. The user can easily add new fields computed on the fly in \idefix VTK outputs to simplify visualisation and post-processing. The VTK outputs produced by \idefix can be directly loaded in Paraview or Visit \citep{Childs_VisIt_An_End-User_2012} for instantaneous visualisation. Python routines are provided with the code to load \idefix VTK and dump files into numpy arrays. Both VTK and dump files can also be analyzed with the \texttt{yt} Python library \citep{Turk.Smith.ea11} with the \texttt{yt\_idefix} extension\footnote{\href{https://pypi.org/project/yt-idefix/}{https://pypi.org/project/yt-idefix/}}.
Note that the VTK file format imposes data be written in single precision and only contains cell-averaged quantities (reconstructed from face-averaged quantities when storing $B$). This reduces storage requirements but also prevents the code from restarting from VTK files, hence the necessity for dump files. 

Internally, \idefix relies on MPI I/O routines to write all of the data from all of the MPI processes to a single file (either VTK or dump). Hence, no external I/O library is required to compile and run \idefix. We found this property useful when moving setups on different machines as it greatly simplifies the portability.  

\subsection{Parallelism \label{sec:mpi}}

When run in serial, \idefix starts a single process on the host and launches the computation kernels (the \verb!idefix_for!) either on a single device (typically a GPU) or in the host process (when running without any accelerator). In addition, \idefix supports coarse-grained parallelism, as it can run on several processors or nodes using domain decomposition and the Message Passing Interface library (MPI).

When MPI parallelisation is enabled, \idefix decomposes the grid into sub-domains of similar size and attaches each MPI host process to a different device. MPI is then used to exchange boundary elements directly between devices, using for instance MPI-Cuda aware implementation on NVIDIA GPUs. The MPI routines have been optimised to pack and unpack boundary elements in device space, and are based on MPI persistent communications to reduce latency. Hence, if a direct connection is available between devices (such as NVlink), there is no host-to-device overhead induced by communications.

The boundary exchange routines in \idefix use a sequential dimension decomposition strategy. This means that we first exchange all of the boundary elements in the first dimension before starting to exchange in the second dimension and so on for the third dimension. This implies that in 3D, the exchange of ghost zones in the 3rd dimension has to wait for the exchanges in the first and second dimensions. This choice was motivated by direct timing measures comparing the sequential implementation to the simultaneous implementation where the face, corner and edge elements are all exchanged at the same time with the 26 neighbours in 3D (see e.g. the implementation in \Athena, \citealt{Stone.Tomida.ea20}). The sequential implementation proved to be generally faster, especially on NVidia GPUs, so we chose it. We note however that such a choice would not be suited for an AMR version of \idefix.

\section{Tests}
\subsection{Hydrodynamical tests}
\paragraph{Hydrodynamical Sod shock tube:}

\begin{figure}
   \centering
   \includegraphics[width=\hsize]{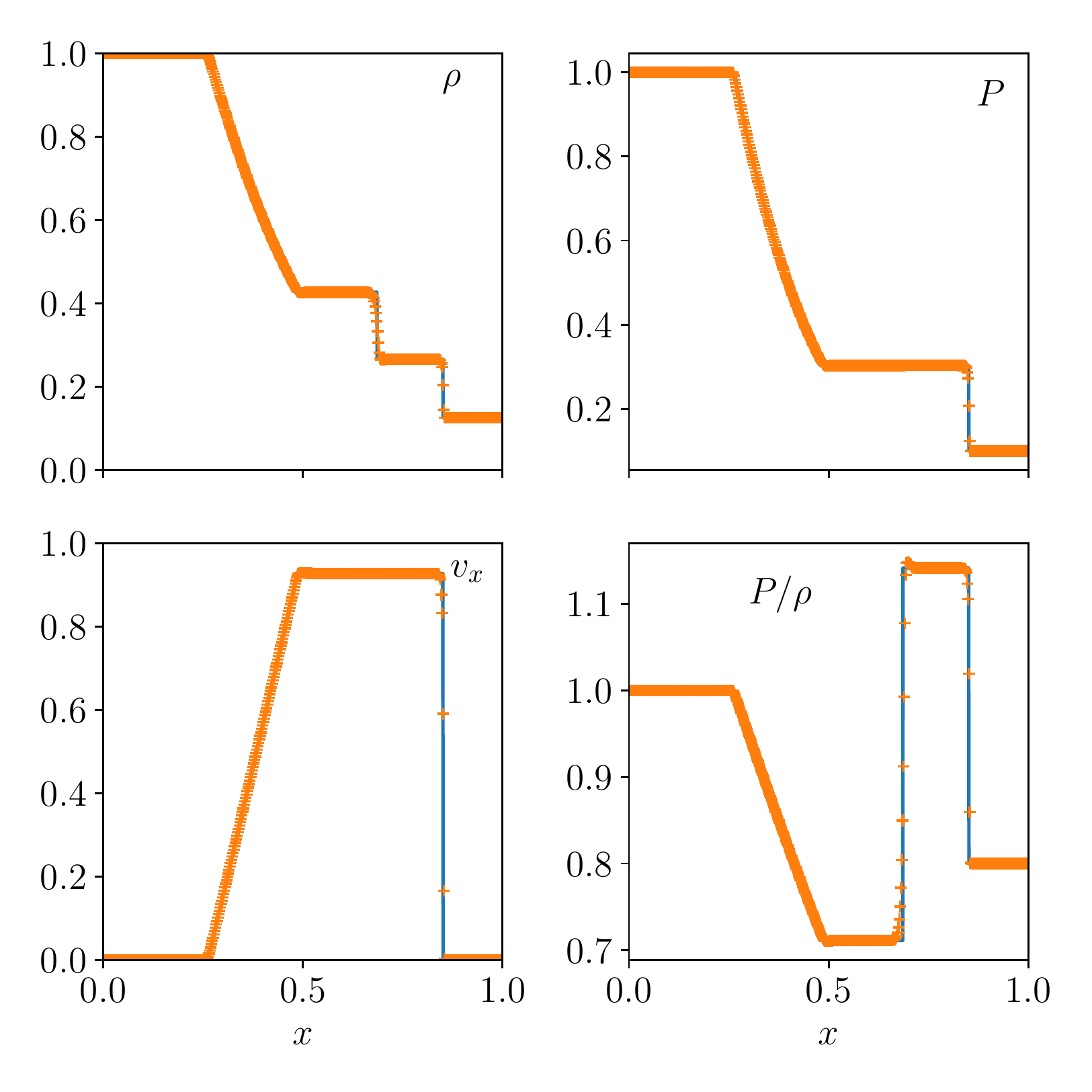}
      \caption{Adiabatic hydrodynamical \cite{Sod78} shock test, at $t=0.2$. Orange crosses: \idefix solution computed using the Roe solver with 500 points. Blue plain line: analytical solution.}
         \label{fig:idefix-sodHD}
   \end{figure}

A standard and simple 1D test for hydrodynamic codes, the Sod shock tube \citep{Sod78} is made of two constant states separated by a discontinuity. Here we use $\rho_L=1.0$, $P_L=0.1$ on the left of the discontinuity and $\rho_R=0.125$, $P_R=0.1$ of the right, with $v=0$ in the entire domain initially. The test is performed with an ideal equation of state having $\gamma=1.4$ and is integrated up to $t=0.2$ using various combinations of Riemann solvers and time integrators. An example with the Roe Riemann solver, linear reconstruction and RK2 time-integrator is reproduced in Fig.~\ref{fig:idefix-sodHD} and compared against the analytical solution.

 \paragraph{Double Mach reflection test:} 
 \begin{figure*}[!ht]
   \centering
   \hspace*{-2cm} \includegraphics[width=1.2\hsize]{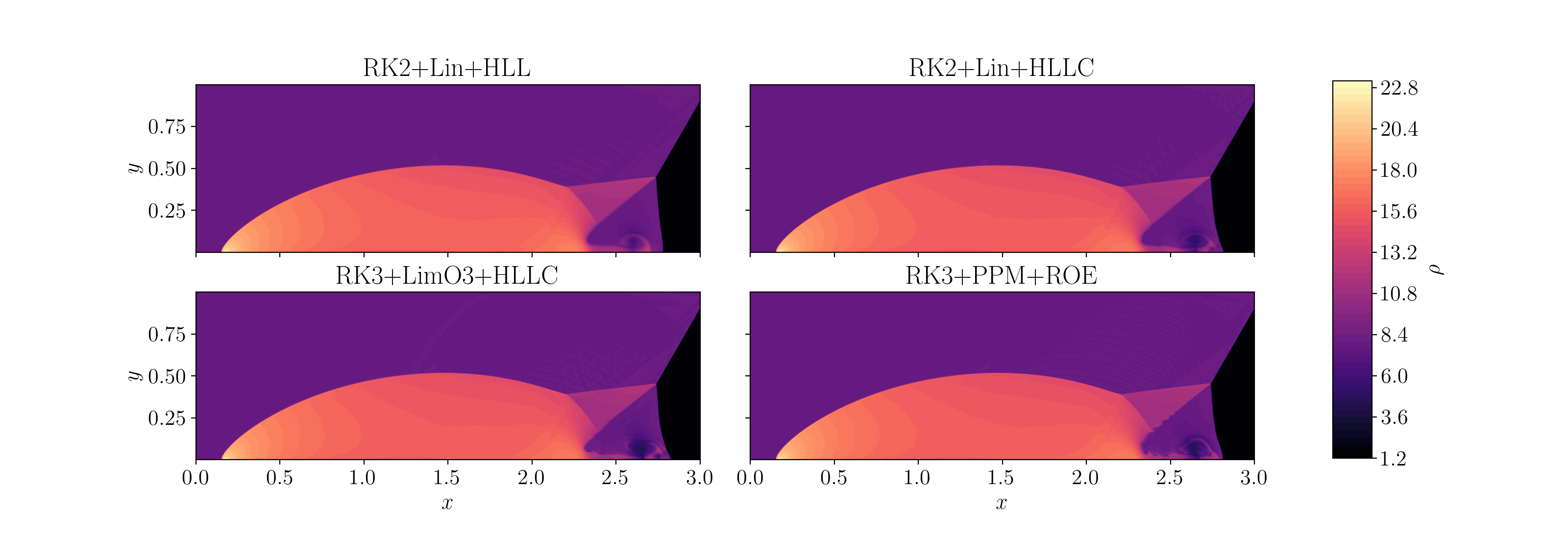}
      \caption{Double Mach reflection test computed at $t=0.2$. We represent here the resulting density map for various combination of algorithms with a $1920\times 480$ resolution. }
         \label{fig:idefix-MachReflection}
   \end{figure*}
   
   \begin{figure*}[!ht]
   \centering
   \hspace*{-2cm} \includegraphics[width=1.2\hsize]{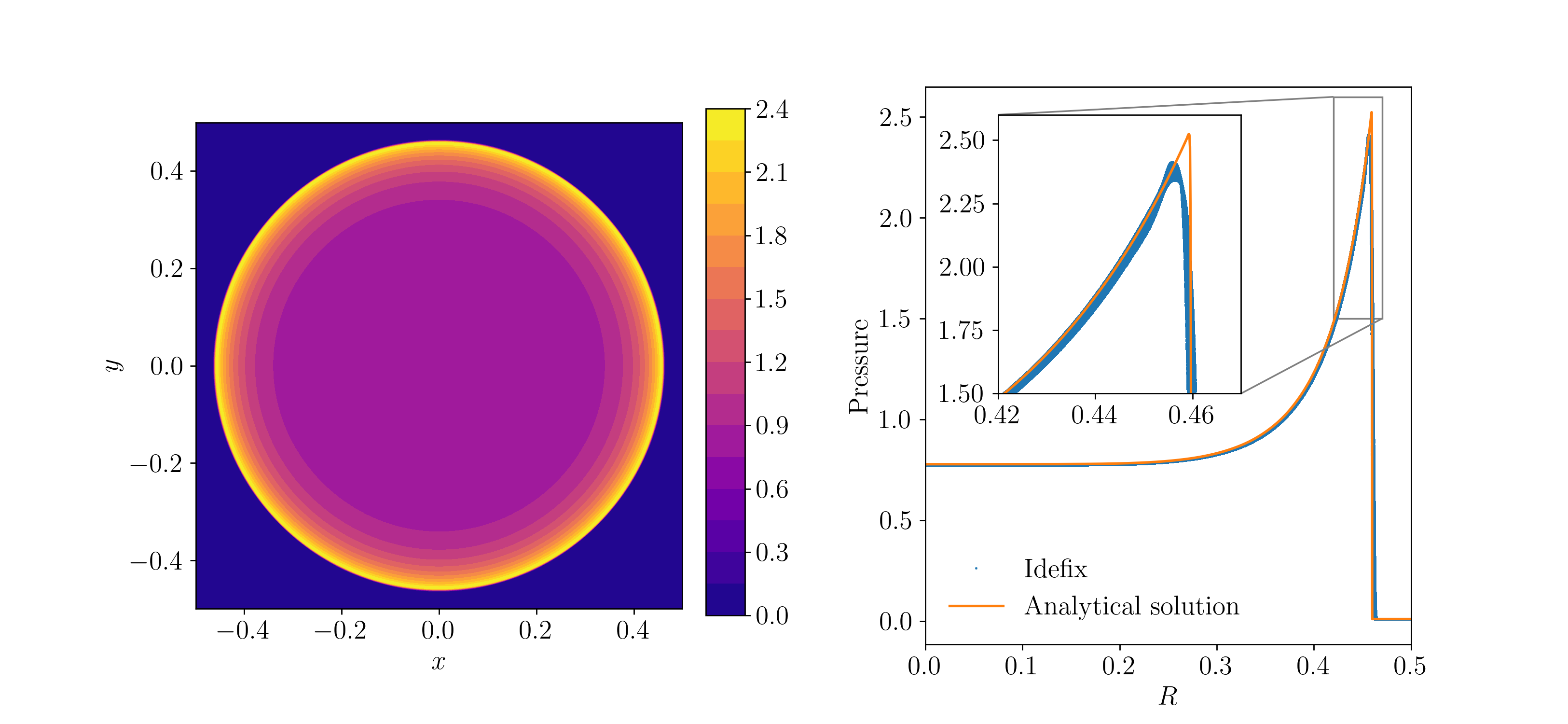}
      \caption{Sedov-Taylor blast wave in cartesian geometry at $t=0.1$. Left panel: slice of the pressure field through $z=0$. Right pannel: pressure distribution as a function of the distance from the blast center. The inset shows a zoom around the blast shock region.}
         \label{fig:idefix-BlastCartesian}
   \end{figure*}
   
    \begin{figure*}[t!]
   \centering
   \hspace*{-2cm} \includegraphics[width=1.2\hsize]{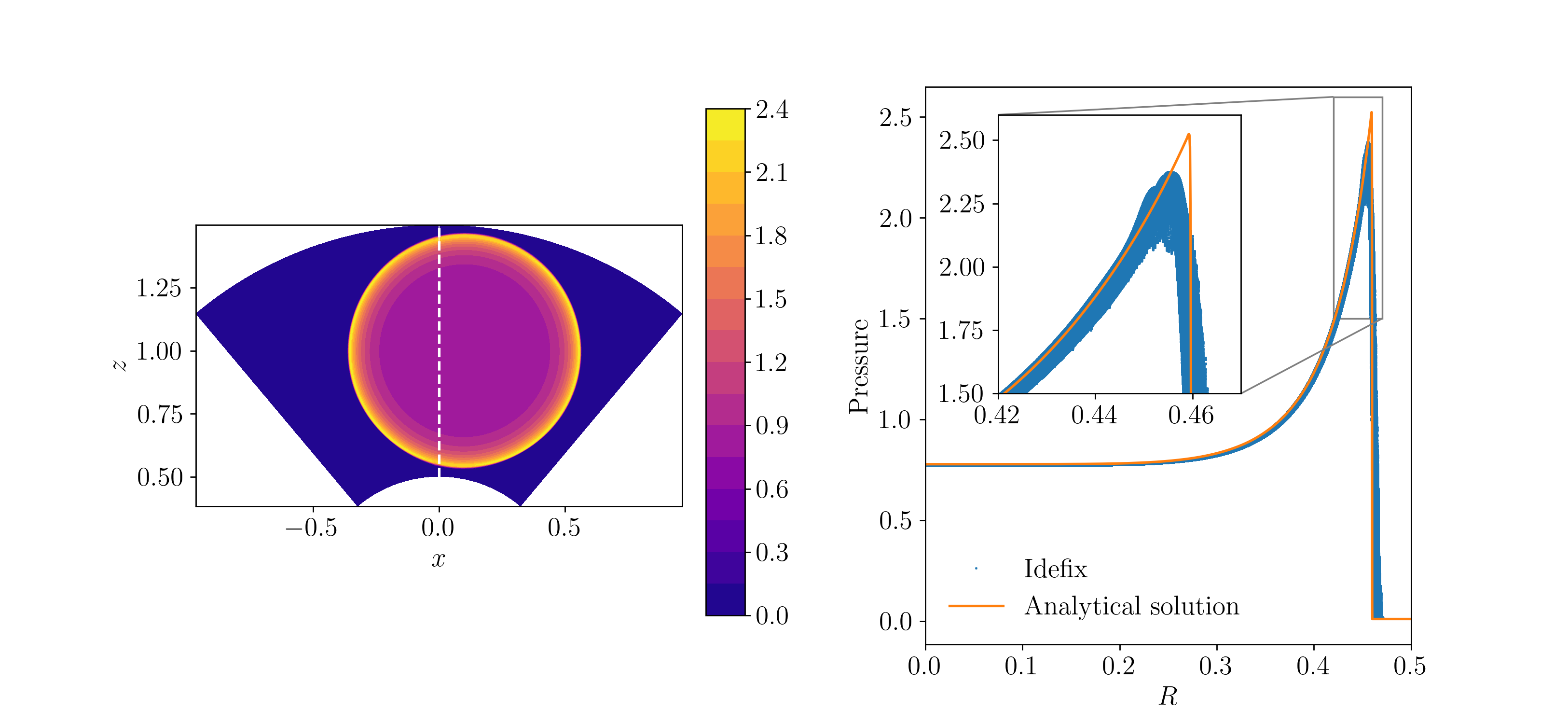}
      \caption{Sedov-Taylor blast wave in spherical geometry at $t=0.1$. The blast is initialised slightly off the polar axis (at $x=0.1,z=1$). Left panel: slice of the pressure field through $y=0$. The dashed white line represent the polar axis of the domain, which is regularised automatically by \idefix. Right panel: pressure distribution as a function of the distance from the blast center. The inset shows a zoom around the blast shock region.}
         \label{fig:idefix-BlastSpherical}
   \end{figure*}

 A test introduced by \cite{Colella.Woodward84} in which the interaction between two shock waves forms a contact discontinuity and a small jet. The properties of this jet are known to depend strongly on numerical diffusion and in particular on the spatial reconstruction scheme and Riemann solver. The test presented here (Fig.~\ref{fig:idefix-MachReflection}) follows the initial conditions from \cite{Colella.Woodward84} and uses various combinations of reconstruction, Riemann solvers and time integrator on a $1920\times 480$ resolution grid. As expected we find a more vigorous jet along with the beginning of Kelvin-Helmholtz rolls for the LimO3 and PPM reconstruction schemes that are the least diffusive schemes in \idefix. This figure can be compared with fig.~16 from \cite{Stone.Gardiner.ea08} and fig.~2 from \cite{Mignone.Bodo.ea07}.

\paragraph{Sedov-Taylor blast wave:} 

   This test aims at reproducing the \cite{Sedov46} and \cite{Taylor50} blast wave self-similar solution for infinitely compact explosions. In this test, we consider an initially stationary flow with $\rho=1$, $P=10^{-2}$ and $\gamma=5/3$. We initiate the explosion in the centre of the domain by depositing an amount of energy $E=\iiint_{r<r_i} P\,\mathrm{d}V/(\gamma-1)=1$ in a spherical subdomain of radius $r_i$. The problem is then integrated up to $t=0.1$ using the HLL solver, RK2 time-integration and linear reconstruction, at which point we compare the numerical solution to the analytical one. We perform 2 tests: one in cartesian geometry with a resolution $512^3$ and a cubic box size $L=1$ (Fig.~\ref{fig:idefix-BlastCartesian}), and one in spherical geometry ($n_r\times n_\theta\times n_\phi=256\times 256\times 512$ with a domain size $r\in[0.5,1.5]$, $\theta\in [0,0.7]$, $\phi\in[0,2\pi]$), where the explosion is initiated close to the polar axis ($x=0.1,\,y=0,\,z=1$) so as to test the propagation of the blast wave through the axis (Fig.~\ref{fig:idefix-BlastSpherical}). In the cartesian test, all three directions are assumed to be periodic, while in the spherical test, we use outflow boundary conditions in radius, the axis regularisation boundary condition in the $\theta$ direction, and periodic boundary conditions in azimuth.

\subsection{MHD tests}
\label{sec:mhd-tests}
 
\paragraph{MHD shock tube:} 
\begin{figure}[t!]
   \centering
   \includegraphics[width=1.1\hsize]{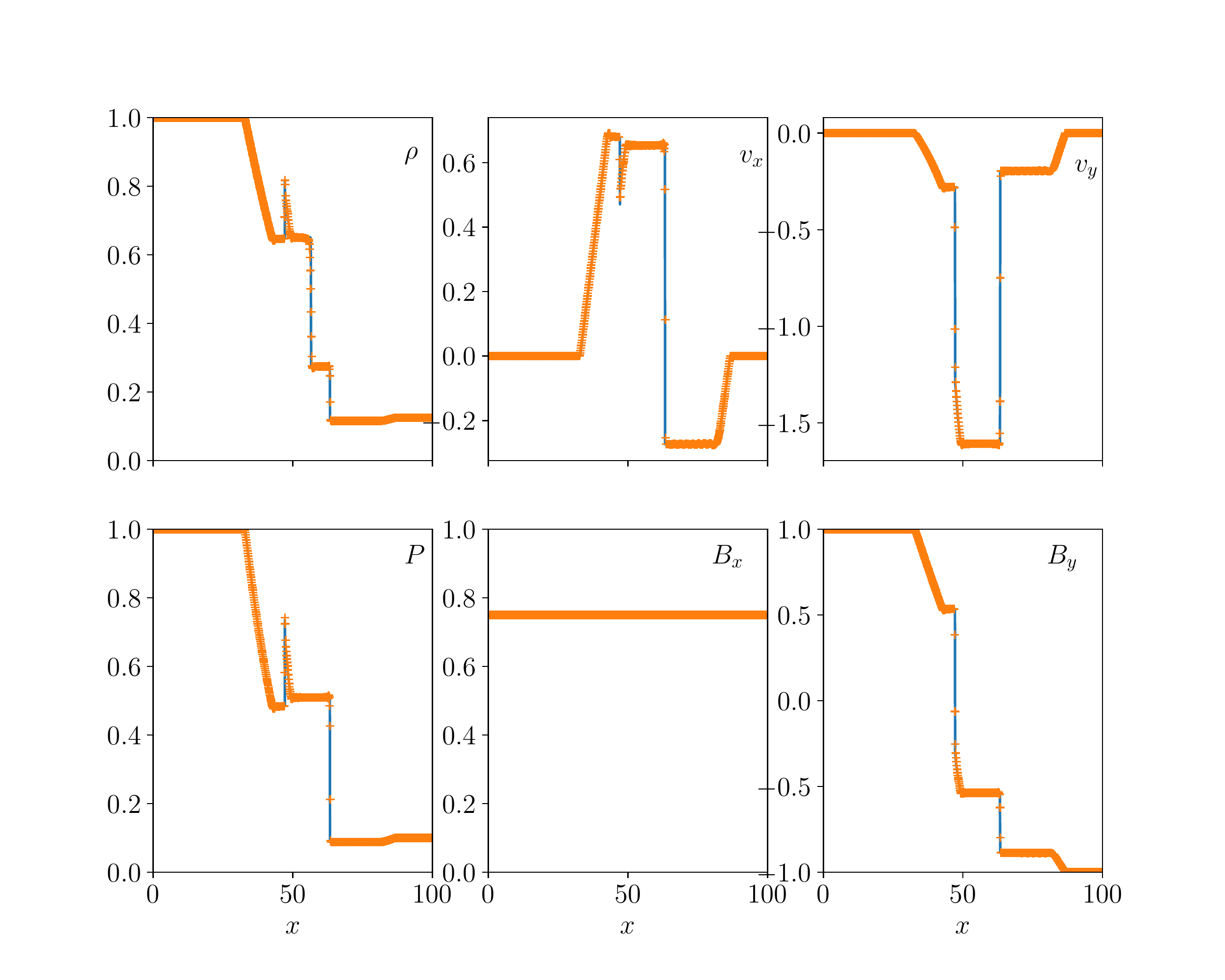}
      \caption{\cite{Brio.Wu88} adiabatic MHD shock test. \idefix solution (orange cross) was computed using the Roe solver with 500 points and linear reconstruction while the reference solution (blue line) was computed with Pluto 4.3 using 8000 points.}
         \label{fig:idefix-sodMHD}
 \end{figure}
 
The \cite{Brio.Wu88} MHD shock tube test is essentially the MHD equivalent of the Sod shock tube test in hydrodynamics. We use $\gamma=5/3$ and an initial condition identical to \cite{Mignone.Bodo.ea07}. There is no analytical solution for this test, so the reference solution is computed with Pluto 4.3 using a Roe solver and a grid of 8000 points (Fig.\ref{fig:idefix-sodMHD}). Note that this test being 1D, it only validates the Riemann solvers but not the EMF reconstruction schemes that are used only in multi-dimensional tests. We observe an excellent agreement with the high resolution solution. We note a small amplitude oscillation of the post-shock velocity that is typical of the primitive variable reconstruction used in \idefix.

\paragraph{Linear MHD wave convergence:} 
\begin{figure*}[t!]
   \centering
    \includegraphics[width=1.0\hsize]{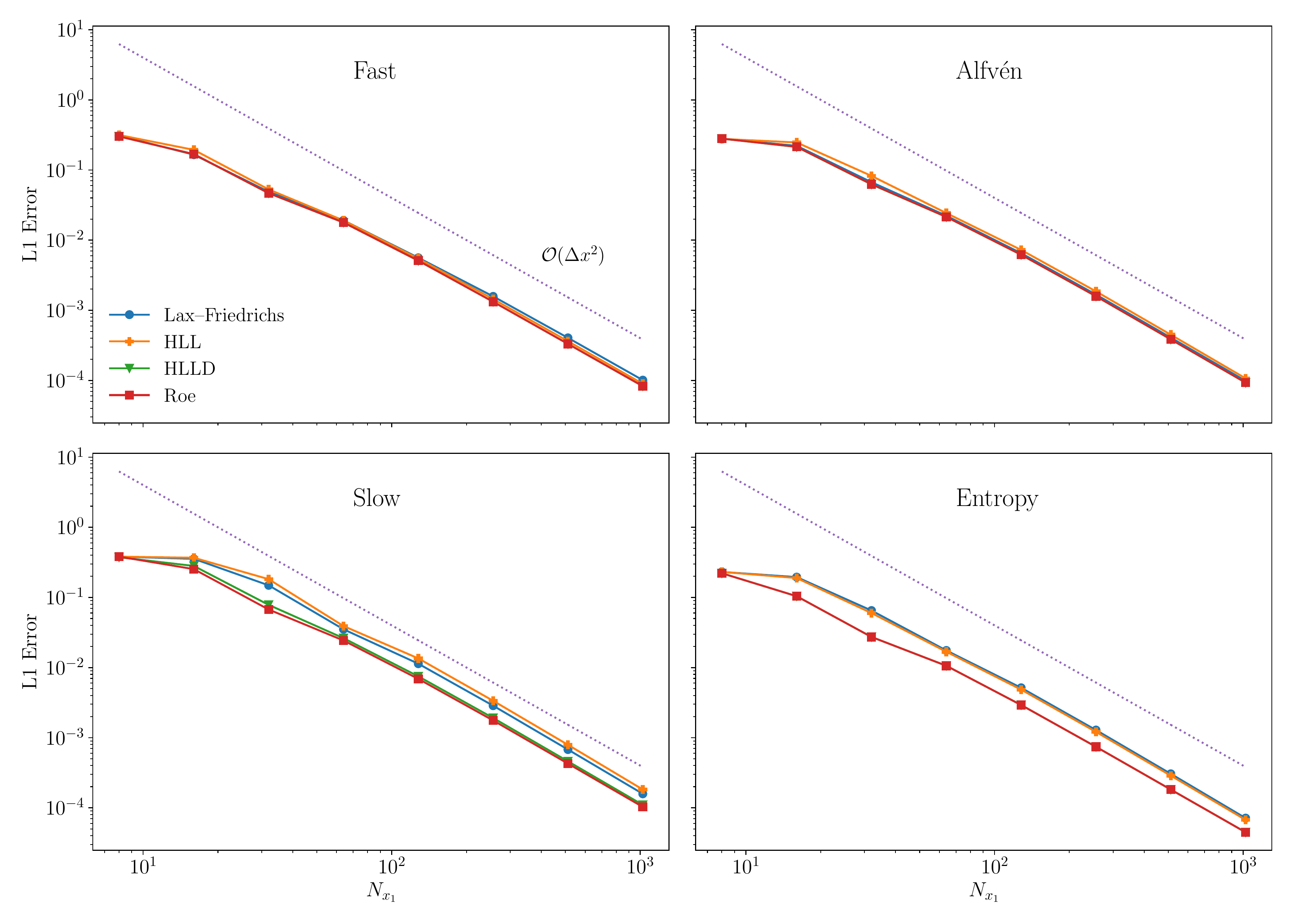}
      \caption{Convergence rate of the 4 linear wave types in adiabatic MHD for various Riemann solvers using linear reconstruction. The code exhibits seconder order spatial convergence, as expected.}
         \label{fig:convergenceLinear}
   \end{figure*}

\begin{figure*}[t!]
   \centering
    \includegraphics[width=1.0\hsize]{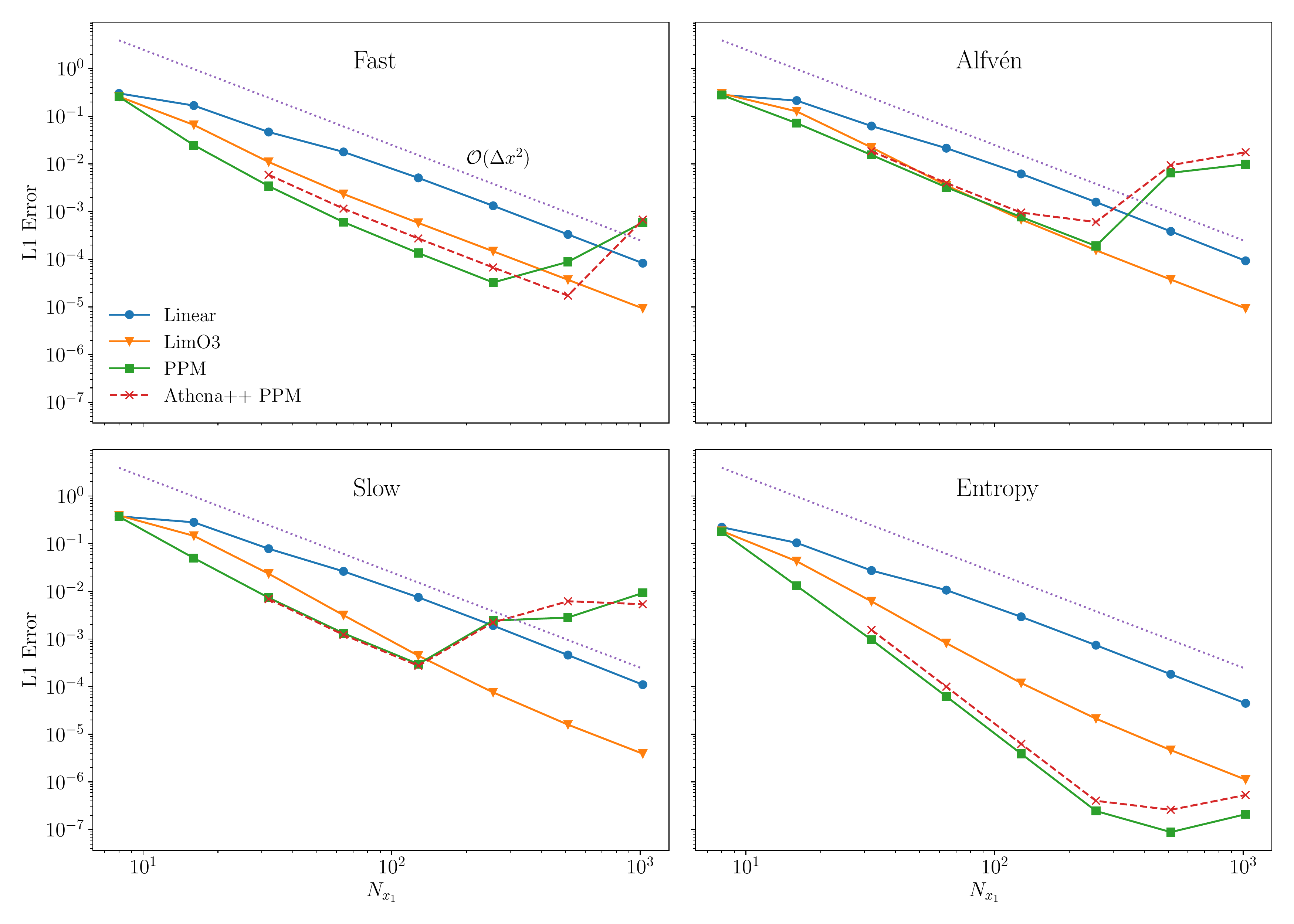}
      \caption{Convergence rate of the 4 linear wave types for various reconstruction techniques using the HLLD Riemann solver. The code exhibits a convergence rate slightly larger than second order for LimO3. Note however the lack of convergence at high resolution for PPM, which is also observed in \Athena{}++ (dashed lines).}.
         \label{fig:convergenceOrder}
   \end{figure*}
   
    \begin{figure*}[h!]
   \centering
   \hspace*{-2cm} \includegraphics[width=0.9\hsize]{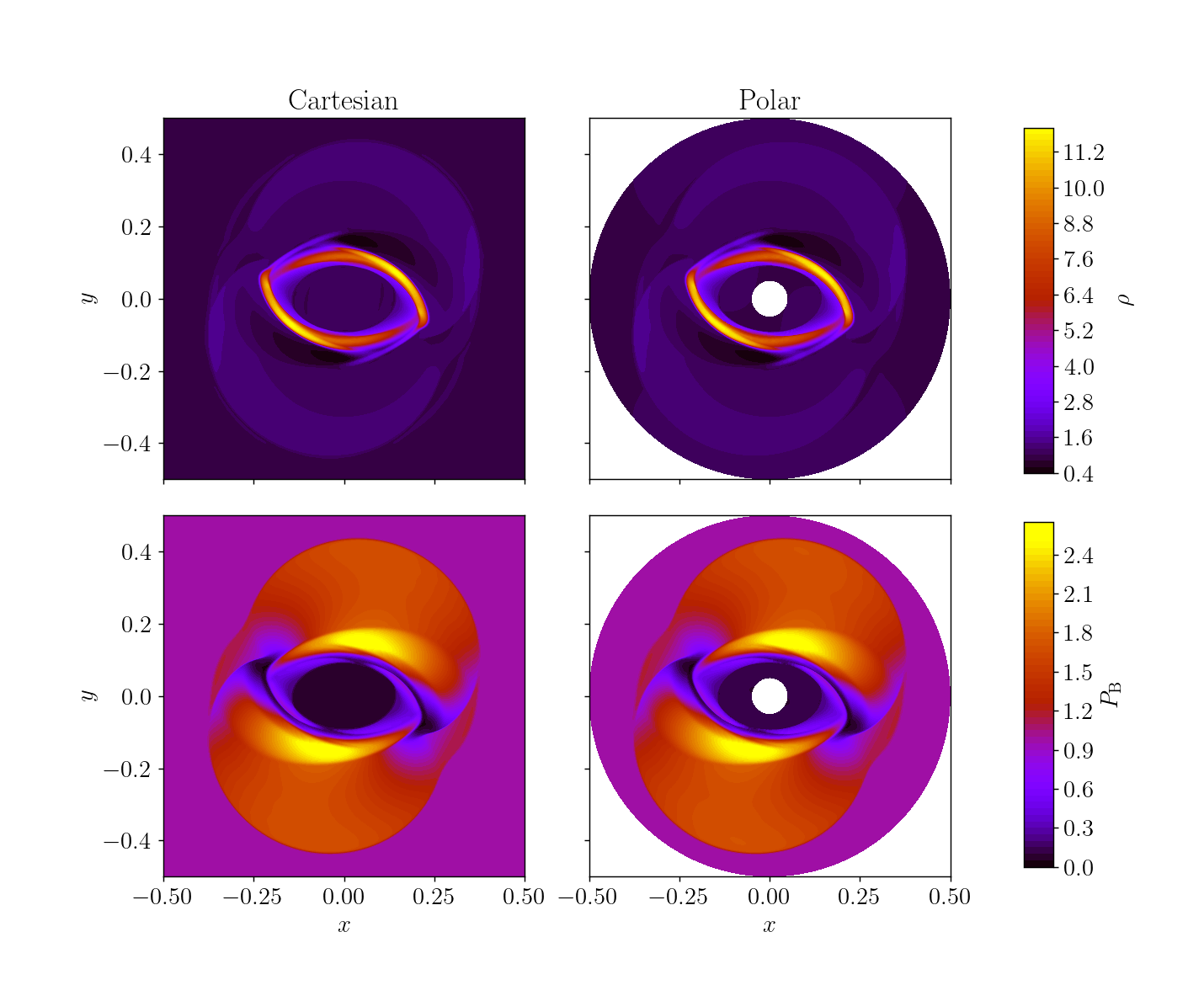}
      \caption{Magnetic rotor test from \cite{Balsara.Spicer99a}, on the left using a cartesian geometry and on the right using a polar geometry. Note the close resemblance of the iso-surface indicating the excellent behaviour of the code in these two geometries.}
         \label{fig:idefix-magRotor}
  \end{figure*}

This test was designed by \cite{Gardiner.Stone05} and \cite{Stone.Gardiner.ea08} to recover the linear MHD eigenmodes on an inclined grid. The test consists of a rectangular box $(L_x,\,L_y\,L_z)=(3.0,1.5,1.5)$ and a grid of $2N\times N\times N$ cells, which is assumed to be periodic with the three directions of space. The background flow is assumed to be stationary (except for the entropy wave) with $\rho=1$ and $P=\gamma^{-1}=3/5$. To fully characterise the initial condition, we use a coordinate system $(x_1,x_2,x_3)$ that is rotated with respect to the grid $(x,y,z)$. In this rotated coordinate system, we use $B_1=1$, $B_2=3/2$ and $v_1=1$ (when testing the entropy mode), otherwise $v_1=0$. With this background state, the fast, Alfv\'en and slow modes propagate respectively at $c_f=2,c_\mathrm{A}=1,\,c_s=0.5$ in the $x_1$ direction, while the entropy mode propagates at $v_1$. To check the convergence of the code, we set up a small amplitude ($\epsilon=10^{-6}$) monochromatic wave along $x_1$ for each eigenmode (see \citealt{Stone.Gardiner.ea08} for the expression of the eigenmodes). We then compare the value of all of the fields after exactly one crossing time $t_c$ of the simulation domain and compute the error $e$ as

\begin{align}
    e = \sqrt{\sum_n\left(\frac{\sum_{ijk}|\delta q_{i,j,k,n}^0-\delta q_{i,j,k,n}^{t_c}|}{\epsilon N_x N_y N_z N_\mathrm{components}}\right)^2}
\end{align}
where $N_\mathrm{components}$ is the number of non-zero components in the eigenmode and $\delta q_{i,j,k,n}^t$ is the deviation from the background state at time $t$ for component $n$ at location $(x_i, y_j, z_k)$. We perform two kinds of linear convergence tests: we first compare the different Riemann solvers with linear reconstruction and second-order time integration (RK2) (Fig.~\ref{fig:convergenceLinear}). This confirms that \idefix is overall second order for smooth problems, as expected. We next perform this test using different reconstruction schemes with the HLLD Riemann solver: linear (van Leer) with RK2 timestepping and LimO3 and PPM in conjonction with RK3 timestepping (Fig.~\ref{fig:convergenceOrder}). This demonstrates the superiority of LimO3 and PPM schemes which overall have a lower diffusivity than linear (particularly notable for the entropy mode). We note however that at high resolution, the PPM reconstruction scheme stops converging for this test. This is particularly stringent for the slow mode, which seems to be most sensitive to this effect (probably because it has the lowest wave speed and hence require longer integration times). We have checked that this lack of convergence was also observed with more diffusive solvers (HLL), and is also present in other codes (notably \Athena{}++ with primitive reconstruction, see figure \ref{fig:convergenceOrder}), and hence is not related to the PPM implementation used in \idefix. The error reached at very high resolution is about $10^{-2}$ for PPM, which is potentially problematic for sensitive applications. Overall, it indicates that the PPM scheme could produce low level numerical artifacts in very high resolution simulations, and that the LimO3 scheme should be preferred, as it shows proper second order convergence for all of the resolutions we have tested.

\paragraph{Magnetised rotor test:} This test is designed to test the propagation of strong torsional Alfv\'en wave \citep{Balsara.Spicer99a}. It consists of a fast-rotating disc (the rotor) embedded in a light stationary fluid (the ambient medium). The whole domain is initially threaded by a homogeneous magnetic field directed in the $x$ direction. We follow \cite{Mignone.Bodo.ea07} setting up the problem, with a smooth transition for the angular velocity and density between the rotor and the ambient medium. In  addition, we use two versions of the test in cartesian and polar geometries, to check the validity of our curvilinear implementation. The resulting tests are presented in figure \ref{fig:idefix-magRotor}. The cartesian test is computed on a $512^2$ grid with a Roe Riemann solver, RK2 time integrator, a CFL parameter set to 0.2 and the $\mathcal{E}_c$ EMF reconstruction scheme. The polar test is computed on a $N_R\times N_\varphi=256\times 1024$ grid extending from $R=0.05$ to $R=0.5$ keeping the other parameters of the solver identical to the cartesian setup. This test confirms that \idefix correctly handles curvilinear coordinates such as polar coordinates.

\paragraph{Orszag Tang vortex:} 
\begin{figure*}[t!]
   \centering
   \includegraphics[width=\hsize]{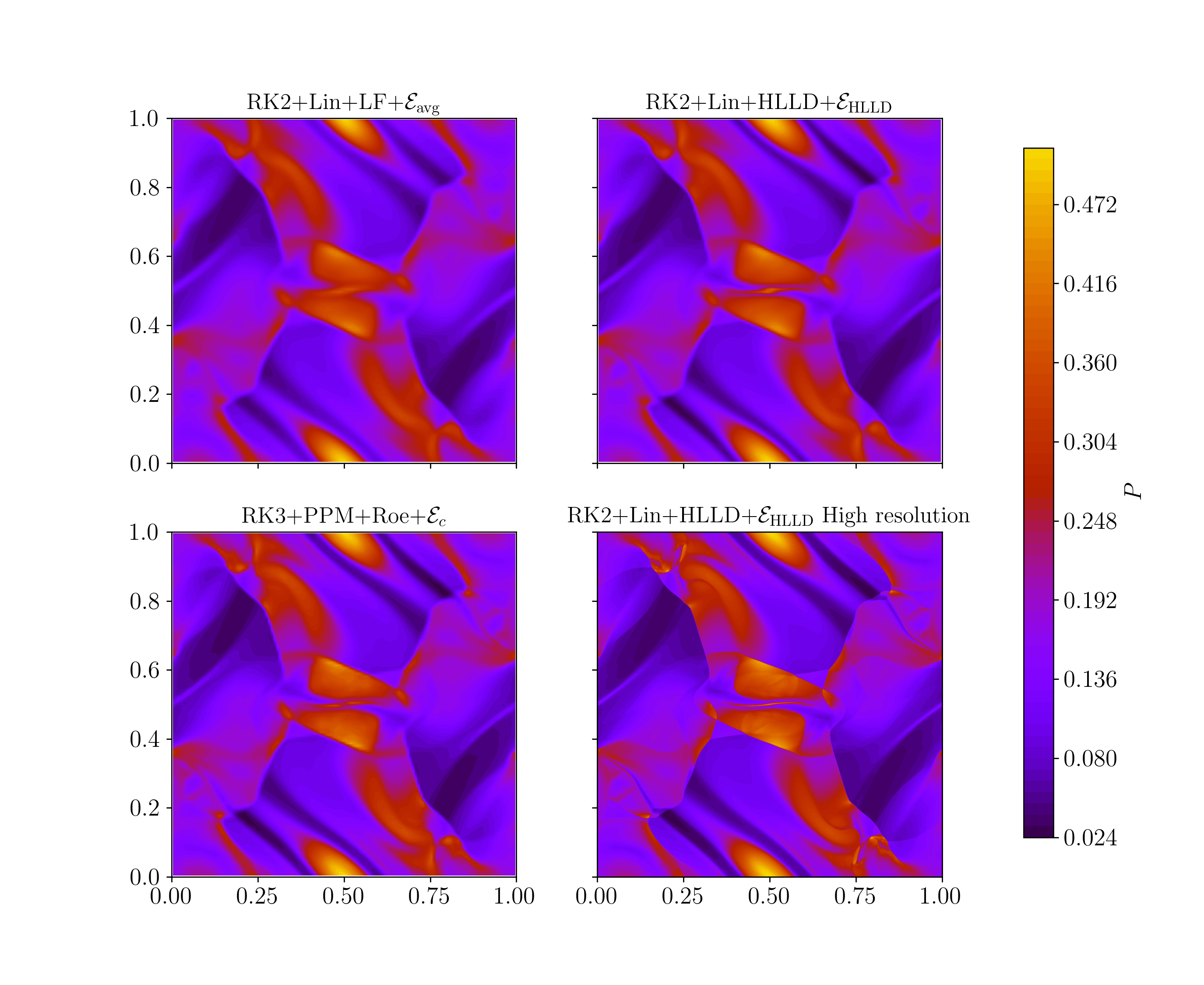}
      \caption{Orszag Tang vortex pressure distribution at $t=0.5$ with different combinations of algorithms computed with $128^2$ points, except for the bottom right plot computed with $1024^2$ points.}
         \label{fig:idefix-OT}
 \end{figure*}
 
 \begin{figure}[t!]
   \centering
   \includegraphics[width=\hsize]{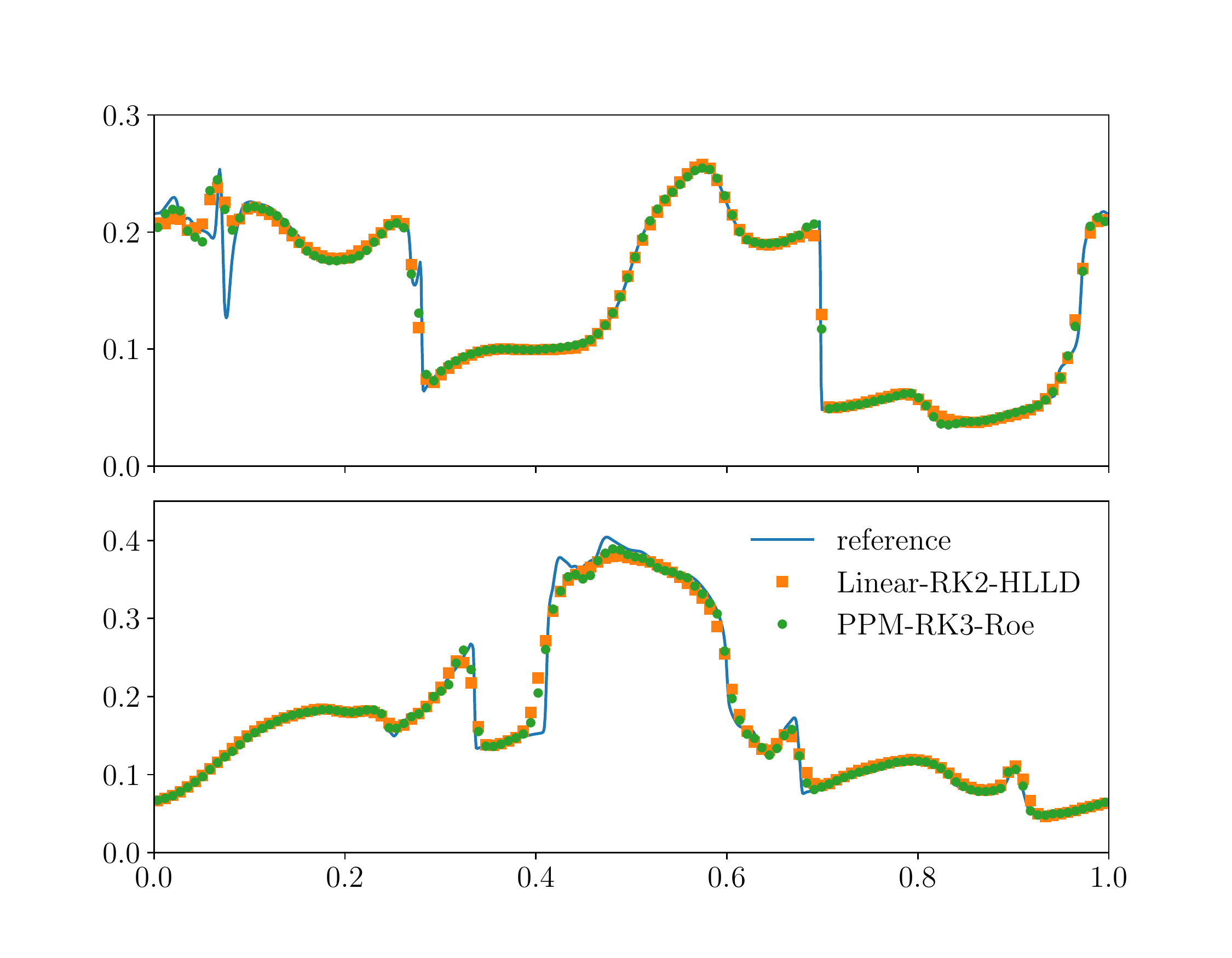}
      \caption{Cut in the Orszag Tang vortex at $y=0.316$ (top) and $y=0.425$ (bottom) for a selected algorithms from figure \ref{fig:idefix-OT}. Solutions were computed with $128^2$ points while the reference was computed with $1024^2$ points}
         \label{fig:idefix-OTcut}
 \end{figure}
 
A classic test of 2D MHD that consists in evolving an initially purely vortical flow threaded by a large scale magnetic field \citep{Orszag.Tang79}. For this test, we use a periodic cartesian domain of size $L_x=L_y=1$ with constant initial pressure $P=5/(12\pi)$ and density $\rho=25/(36\pi)$. We add a velocity perturbation $\bm{v}=-\sin(2\pi y) \bm{e}_x+\sin(2\pi x)\bm{e_y}$ and a magnetic perturbation $\bm{B}=-B_0\sin(2\pi y)\bm{e}_x+B_0\sin(4\pi x)\bm{e_y}$ with $B_0=1/\sqrt{4\pi}$, both of which are initially divergence-free. We then evolve this initial condition up to $t=0.5$ with a resolution $128^3$, plus a reference case at $512^3$. Figure~\ref{fig:idefix-OT} present the resulting pressure distribution using a variety of combination of Riemann solvers, reconstruction scheme and EMF implementations, while figure~\ref{fig:idefix-OTcut} shows a cut at $y=0.316$ and $y=0.425$ of the same pressure field. These figures confirm that the RK3+PPM+Roe combination is the most accurate solver for a given resolution, but other combinations also give the correct results at a reduced numerical cost. This test also validates our implementation of constrained transport since we measure $|\bm{\nabla\cdot B}|\ll 10^{-12}$ in all of these simulations.

\subsection{Additional modules}

\begin{figure*}[h!]
   \centering
    \includegraphics[width=0.9\hsize]{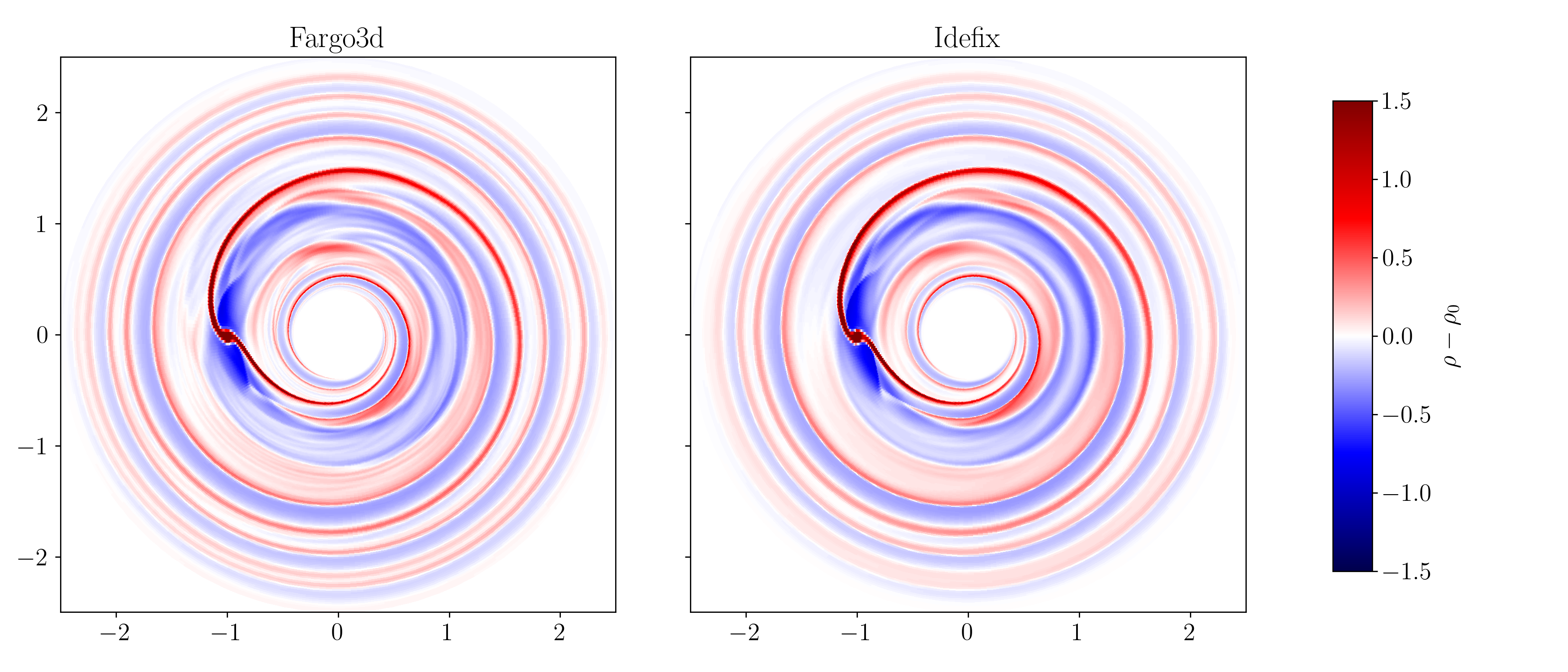}
      \caption{Planet-disc interaction problem with orbital advection enabled at $t=4$ planetary orbits. Comparison between \FargoTD and \idefix for the density deviation map in the planet-disc interaction problem with orbital advection enabled at $t=4$ planetary orbits. Note the presence of small wave-like patterns in \FargoTD that are absent from \idefix .}
         \label{fig:fargo-map}
   \end{figure*}

\begin{figure}[h!]
   \centering
    \includegraphics[width=0.9\hsize]{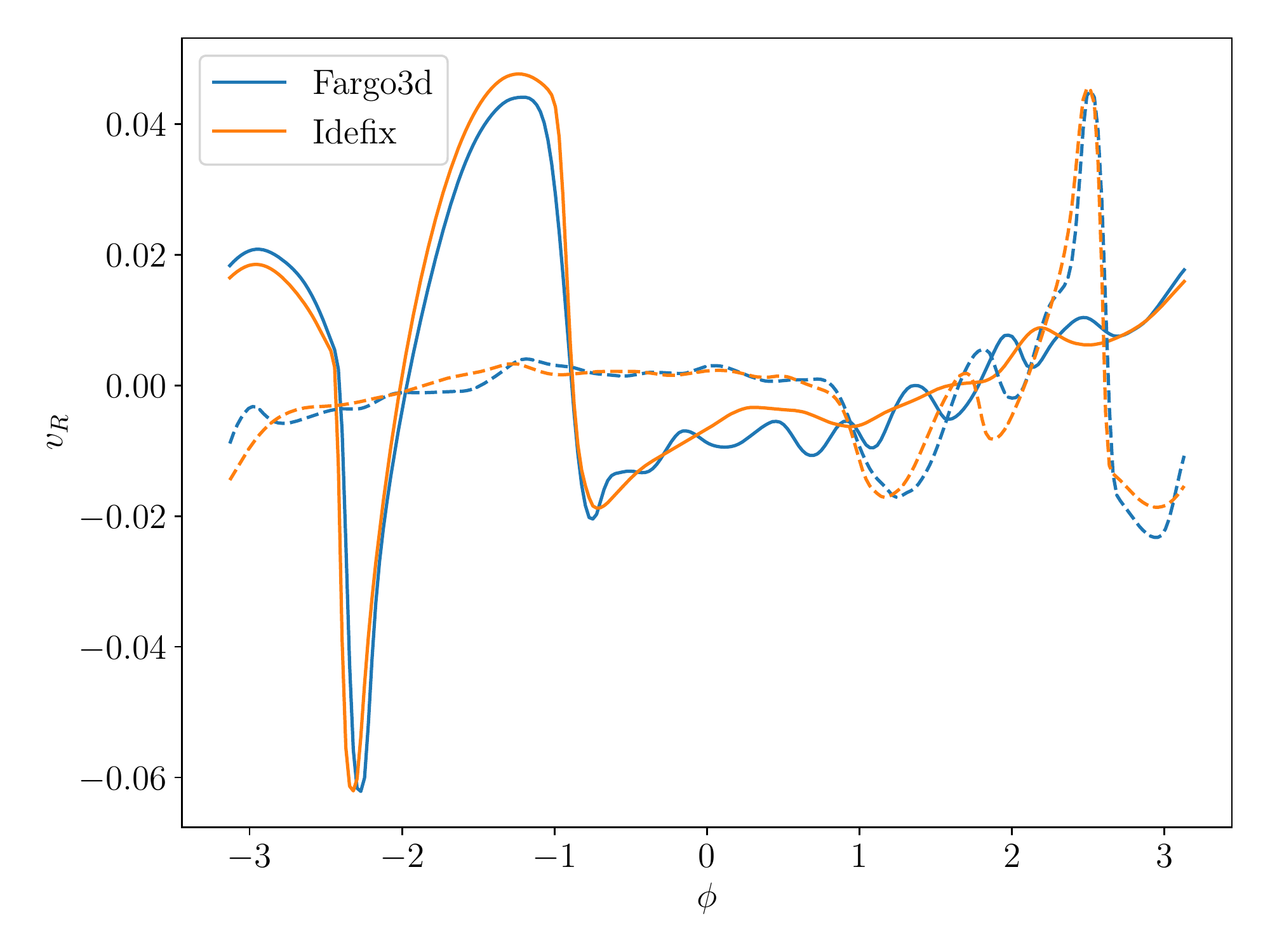}
      \caption{Planet-disc interaction problem with orbital advection enabled at $t=4$ planetary orbits. Comparison between \FargoTD and \idefix of the radial velocity $v_R$ at $R=0.7$ (plain line) and $R=1.3$ (dashed line).}
         \label{fig:fargo-cut}
   \end{figure}

\paragraph{Orbital advection (Fargo-type) module}: We test the orbital advection scheme in hydrodynamics by looking at the perturbation of an embedded planet in an inviscid gaseous disc, solved in 2D polar coordinates. In this test, we set a planet with a planet-to-star mass ratio $q=10^{-3}$, on a fixed circular orbit at $R=1$ with a Plummer potential $ \psi_\mathrm{p}=-q/\big(\lVert\bm{r}-\bm{r}_\mathrm{p}\rVert^2+a^2\big)^{1/2}$ where $a=0.03$ is the Plummer smoothing length. The resolution is set to $(N_R,N_\phi)=(256,256)$ and we use the HLLC Riemann solver with linear reconstruction. Orbital advection is set to use a PPM reconstruction scheme (note that a linear scheme is also available). The problem is then integrated for 4 planetary orbits. We compare the resulting density deviations (Fig.~\ref{fig:fargo-map}) and the radial velocity along the azimuthal direction at several radii (Fig.~\ref{fig:fargo-cut}) using \FargoTD \citep{Benitez.Masset16} as a reference. The agreement between the two codes is excellent. We notice that \FargoTD produces more small-scale waves between the main spiral shocks launched from the planet. 

\begin{figure*}[t!]
   \centering
    \includegraphics[width=0.8\hsize]{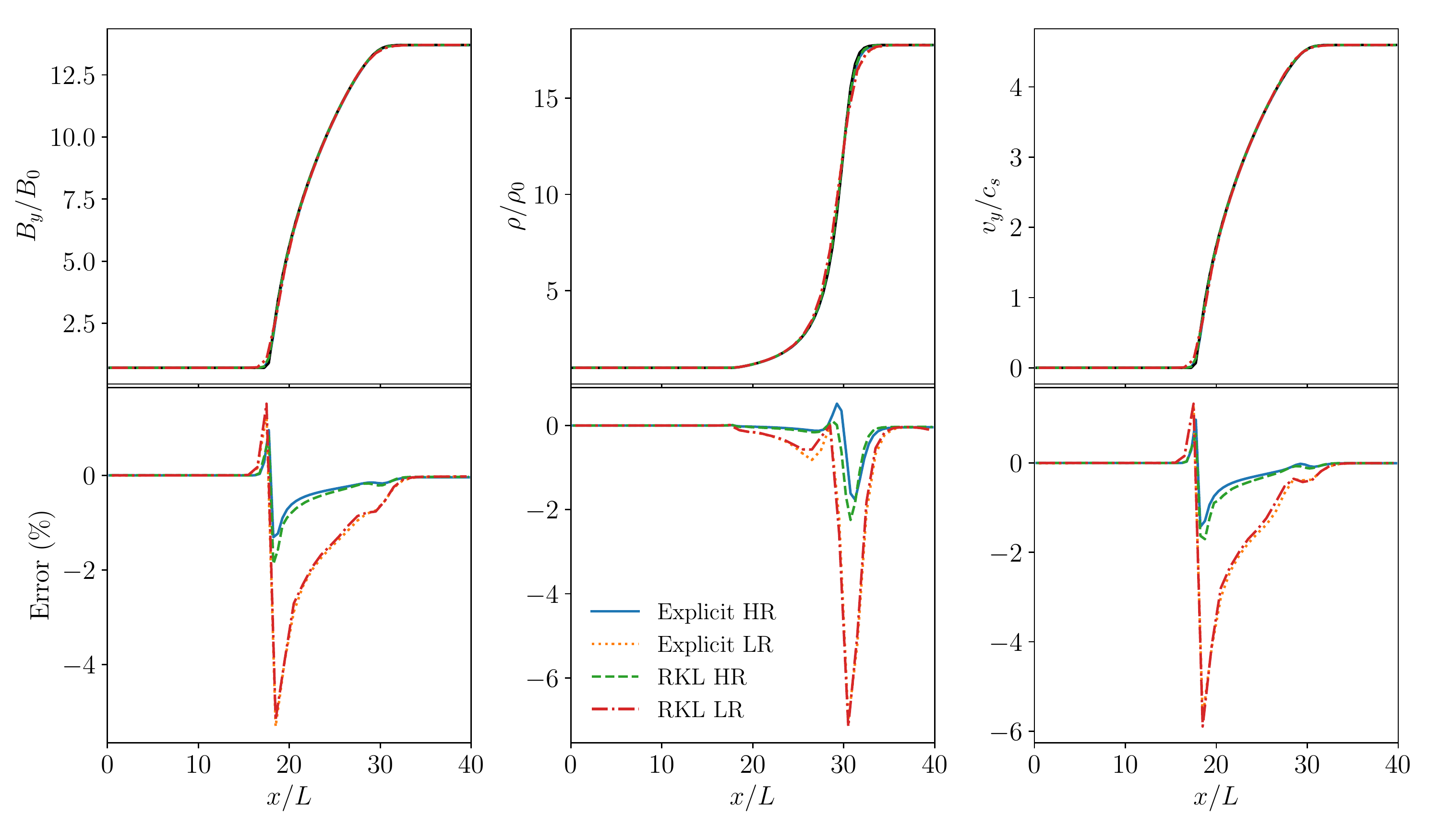}
      \caption{Ambipolar isothermal C-shock integrated at high resolution (HR, 2 points/$L$) and low resolution (LR, 1 point/$L$) and using the explicit or RKL 2nd order time-integrators for parabolic terms. Top: Magnetic field, density and velocity field across the shock (the black line denotes the analytical solution). Bottom: L1 error (in \%) for each field.}
         \label{fig:cshock}
   \end{figure*}

\paragraph{Ambipolar diffusion isothermal C-shock}: This test consists of a ``shock'' (which is actually continuous) and is driven by ambipolar diffusion. We follow \cite{MacLow.Norman.ea95} and use a stationary shock with an upstream Mach number $M=50$, an Alfv\'enic Mach Number $A=10$ and an inclination angle $\theta=\pi/4$ (Fig.~\ref{fig:cshock}). The characteristic length of the shock is $L\equiv\eta_A/v_\mathrm{A}$ where $\eta_\mathrm{AD}\equiv x_\mathrm{AD}B^2$ is the ambipolar diffusivity and $v_\mathrm{A}$ is the Alfv\'en speed. We find an excellent agreement between the analytical solution (in black) and the various solutions computed with \idefix, using either the explicit RK2 integrator or the second order RKL time-integrator for parabolic terms, with an error of less than $2\%$ when using 2 points per $L$. Also, the time to solution with the RKL scheme is about 10\% of the explicit scheme in the high-resolution runs, indicating a very substantial gain brought by the RKL scheme.

        \begin{figure}[t!]
   \centering
    \includegraphics[width=1.0\hsize]{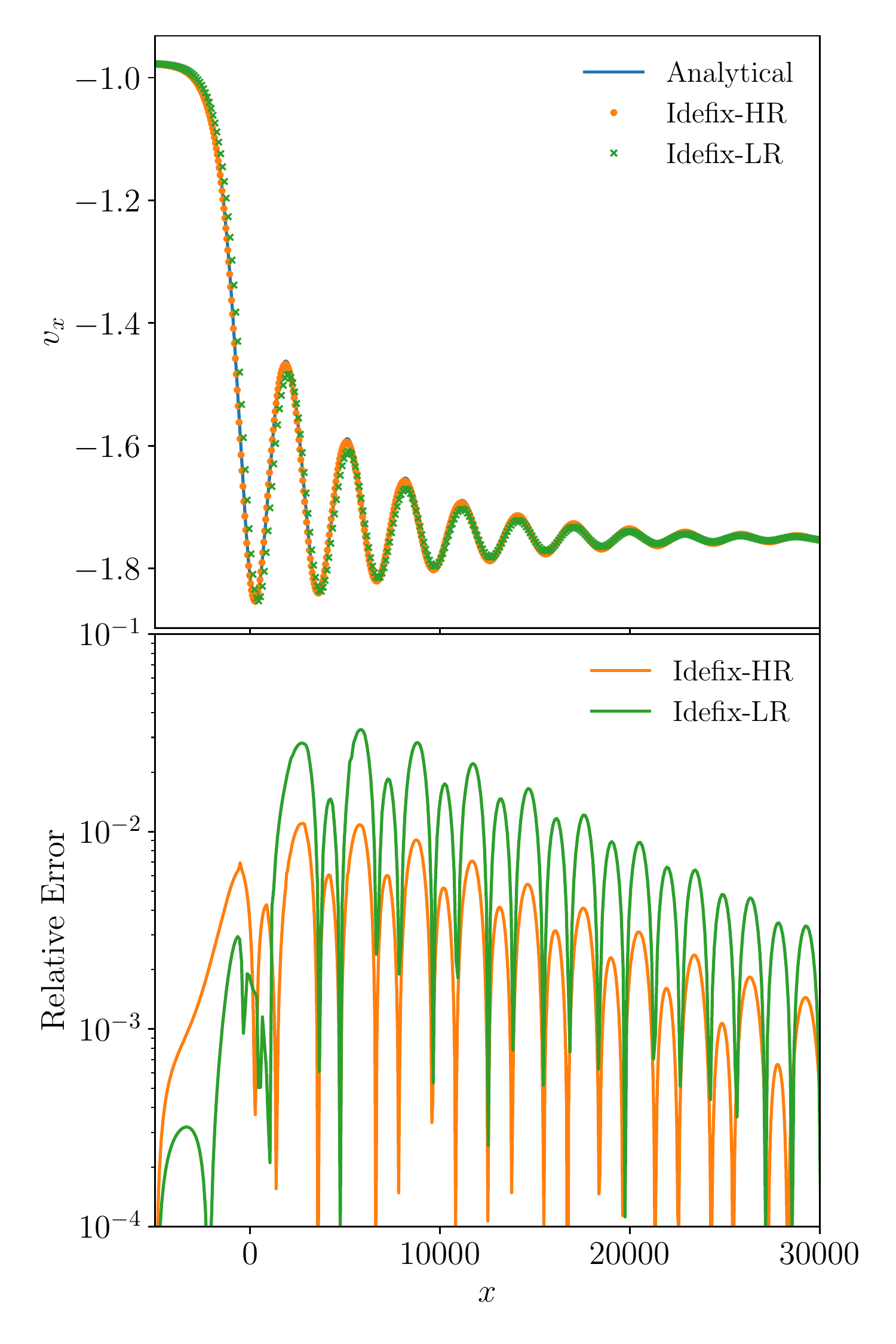}
      \caption{Hall-dominated stationary shock at two different resolutions (the high-resolution test has twice the number of points of the low-resolution test). Top: normal velocity. The propagation of whistler waves in the pre-shock region (right state) is evident. Bottom: relative error between the numerical solution and the semi-analytical one. We observe a convergence rate slightly lower than second order in this case. }
         \label{fig:hallShock}
   \end{figure}
  \paragraph{Hall-dominated shock}: This test consists of a shock wave in a flow dominated by the Hall effect, but also involving Ohmic and ambipolar diffusivities. It was first introduced in the context of multi-fluid MHD by \cite{Falle03} to test the propagation of whistler waves in the pre-shock region of a MHD shock. Here, we follow \cite{Gonzalez-Morales.Khomenko.ea18b} (their shock tube test A\footnote{Their left and right states are not exactly similar to ours since theirs do not satisfy mass conservation initially}) by setting up an initial discontinuity at $x=0$ with the left and right states in table.~\ref{tab:hall-tube}. The system is then evolved with all the non-ideal MHD effects turned on explicitly and with $\eta=1$, $x_A=25$ and $x_H=500$. The solver is set to RK2, CFL to 0.5 and we use the HLL Riemann solver (our only solver compatible with the Hall effect). We finally compare the final steady state obtained to a semi-analytical solution obtained by computing the steady state numerically integrating the induction equations (eq. 50a-50b in \citealt{Gonzalez-Morales.Khomenko.ea18b} \footnote{Note that there is a typo in the last term of eq.~50a in \cite{Gonzalez-Morales.Khomenko.ea18b}, one should read $v_zB_y$ instead of $v_zB_x$}). The results are presented in Fig.~\ref{fig:hallShock}. We find the solution computed with \idefix is in excellent agreement with the analytical solution. The relative error decreases by a factor of 3-4 by doubling the resolution, which indicates a convergence rate slightly less than second order for this test, due to the high diffusivity induced by grid-scale whistler waves. Note that it is possible to strictly recover the second-order convergence using the LimO3 or PPM reconstruction schemes (not shown).
  
  \begin{table}[]
      \centering
      \begin{tabular}{c|c|c}
        Field   & left state & right state \\
        \hline
         $\rho$ & 1.7942 &  1 \\
         $v_x$  & -0.9759 & -1.751 \\
         $v_y$  & -0.6561 & 0\\
         $v_z$  & 0       & 0\\
         $B_x$ & 1        & 1\\
         $B_y$ &  1.74885 & 0.6 \\
         $B_z$ & 0 & 0\\
      \end{tabular}
\caption{Left and right states for the Hall-dominated shock tube. Inspired from \cite{Gonzalez-Morales.Khomenko.ea18b}}
      \label{tab:hall-tube}
  \end{table}

\paragraph{Shearing box}: The shearing box is a well-known model of sheared flows which has been used extensively for the study of the magneto-rotational instability (MRI) in the context of accretion disc physics \citep{Balbus.Hawley91,Hawley.Gammie.ea95}. We have implemented the shearing box specific shear-periodic boundary conditions in \idefix, both in hydro and MHD. This module is also compatible with the orbital advection module, allowing for a significant speedup in large boxes, where the shear flow is largely supersonic. To test the implementation, we compare against the exact shearing wave solution of \cite{Balbus.Hawley06} in hydro, and against a generalized version of it for the MHD case. The setup consists of a cubic box of size $1$ with a Keplerian rotation profile, and a resolution of $256^3$. The initial configuration for the shearing wave is $u_R\equiv u_x=A_R\sin[2\pi(n_R x+n_\phi y + n_z z)]$, where we choose $n_R=0$, $n_\phi=1$, $n_z=4$ and an initial wave amplitude $A_R=10^{-5}$. Additionally, we set a uniform density and pressure $\rho=1$ and $P=1/\gamma$ with $\gamma=5/3$ assuming an ideal equation of state for the gas. Finally, in the MHD case, we add a mean toroidal and vertical magnetic field $B_{y0}=0.02$ and $B_{z0}=0.05$.  Note that while the single shearing wave of \cite{Balbus.Hawley06} is an exact nonlinear solution of the incompressible equations of motion, it is not an exact nonlinear solution in the compressible regime modeled by \idefix. Hence, we only recover the \cite{Balbus.Hawley06} solutions in the linear limit $A_R\ll 1$. 

\begin{figure}[t!]
   \centering
    \includegraphics[width=1.0\hsize]{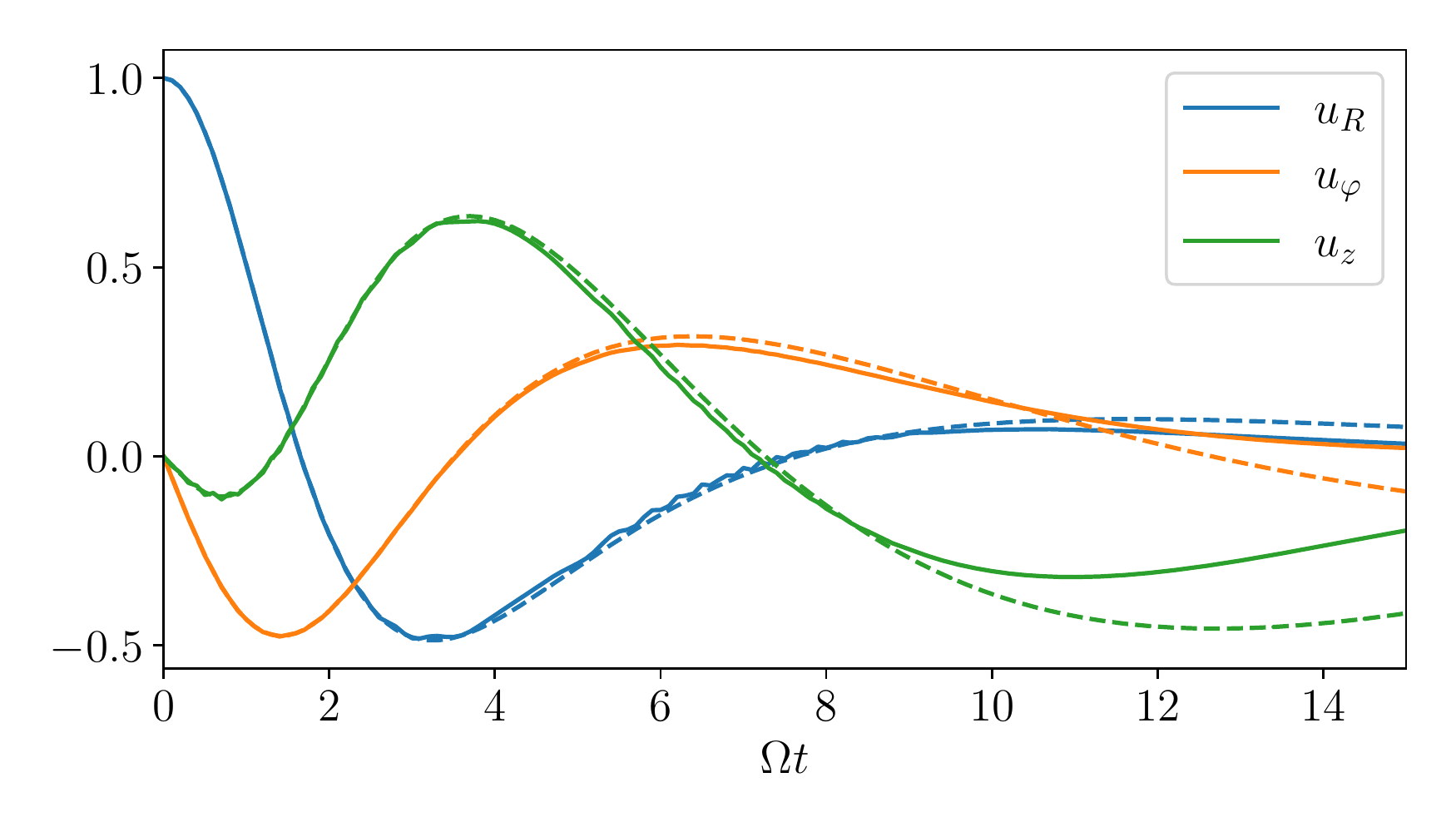}
      \caption{Temporal evolution of the velocity components of the hydro Keplerian shearing wave problem of \cite{Balbus.Hawley06}. The analytical solution is in dashed line, while the \idefix solution computed with $256^3$ points is shown in plain line.  }
         \label{fig:shearingboxHD}
   \end{figure}

\begin{figure}[t!]
   \centering
    \includegraphics[width=1.0\hsize]{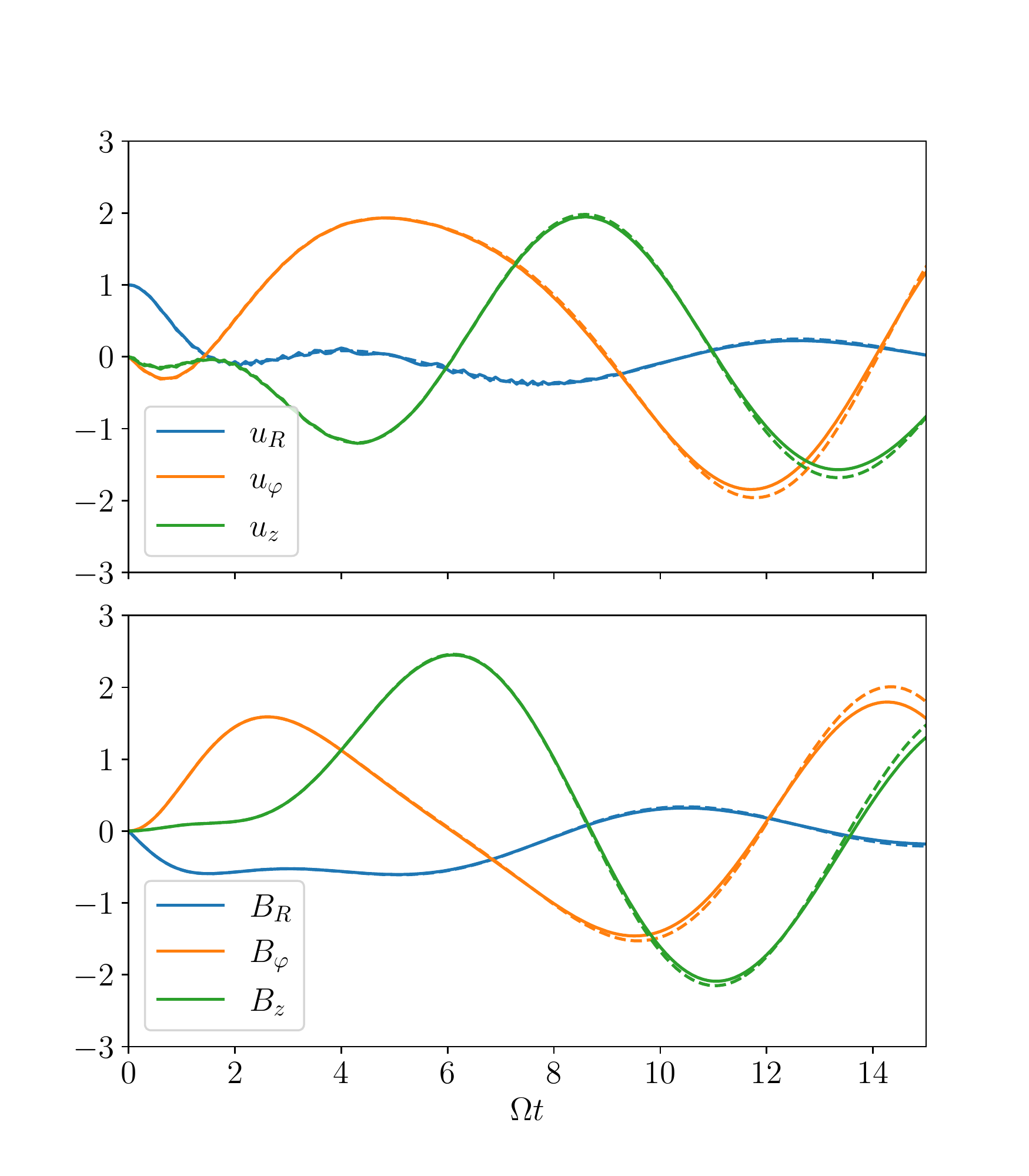}
      \caption{Temporal evolution of the velocity (top) and magnetic (bottom) components of the MHD Keplerian shearing wave problem of \cite{Balbus.Hawley06}. The analytical solution is in dashed line, while the \idefix solution computed with $256^3$ points is shown in plain line.}
         \label{fig:shearingboxMHD}
   \end{figure}

The system is then integrated using the HLLC (hydro) or HLLD (MHD) Riemann solvers using a linear reconstruction, RK2 time-stepping and enabling orbital advection. We present the evolution of the velocity components of the hydro problem in fig.~\ref{fig:shearingboxHD} and the velocity and field components of the MHD problem in fig.~\ref{fig:shearingboxMHD}. The semi-analytical solutions are computed numerically by integrating the linearized equations of motion (eq. 6.25-6.29 in \citealt{Lesur21}). We find that the agreement between the numerical and analytical solution is excellent up to $\Omega t\simeq 10$, where $v_z$ starts to diverge in the hydro case. This time corresponds to an effective radial wavenumber $n_r(t)=3/2\Omega t\sim  15$, so that the wave is effectively resolved by about 17 points. Similar results were obtained with the \Athena code using PPM reconstruction \citep{Balbus.Hawley06}, so they are not necessarily surprising, but they illustrate the difficulty of following shearing waves on long timescales, even at relatively high resolutions.

\section{Performance}

 In order to test the performances of the code on CPUs and on GPUs, we have used an extension of the Orszag Tang problem in 3 dimensions. We consider a periodic cartesian domain of size $L_x=L_y=L_z=1$ with constant initial pressure $P=5/(12\pi)$ and density $\rho=25/(36\pi)$. We add a velocity perturbation $\bm{v}=-\sin(2\pi y) \bm{e}_x+[\sin(2\pi x)+\cos(2\pi z)]\bm{e_y}+\cos(2\pi x)\bm{e_z}$ and a magnetic perturbation $\bm{B}=-B_0\sin(2\pi y)\bm{e}_x+B_0\sin(4\pi x)\bm{e_y}+B_0[\cos(2\pi x)+\sin(2\pi y)]\bm{e_z}$ with $B_0=1/\sqrt{4\pi}$, both of which are initially divergence free. This physical setup is run until $t=1.0$ and the result is compared to that obtained with \pluto to check for convergence.
 The test is executed for both codes with the HLLD Riemann solver, second order (linear) reconstruction, the $\mathcal{E}^c$ EMF reconstruction scheme and RK2 time-stepping in double precision.
 
 We ran this test problem on Intel Cascade lake CPUs and NVidia V100 GPUs on the Jean Zay cluster hosted by IDRIS (CNRS), on AMD Rome CPUs on the Irene Rome cluster hosted by TGCC (CEA), and on AMD Mi250 GPUs on the Adastra cluster hosted by CINES. The details of the configuration and compilation options are given in Tab.~\ref{tab:config}. Note that on CPUs we use the Intel compiler, which allows us to benefit from auto-vectorization. An analysis of the executable produced with Intel Vtune shows that more than 86\% of integration loops are being automatically vectorized with avx2 instructions.  
 
 \begin{table*}[]
    \centering
        \caption{Configuration of the clusters used for the performance tests, and compilation options used to compile \idefix{}.}
    \begin{tabular}{|c|c|c|c|c|c|}
    \hline
       Cluster              & Node configuration & Compiler/Libraries & Compiler option  \\
       \hline\hline
       \makecell{Irene Rome\\TGCC}     & \makecell{2$\times$AMD EPYC 7H12\\(128 cores at 2.6\,GHz) }      & \makecell{Intel 2020.4\\OpenMPI 4.1.4} & -O3 -avx2 -std=gnu++17   \\
       \hline
       \makecell{Jean Zay CSL\\IDRIS}  & \makecell{2$\times$Intel Cascade Lake 6248 \\ (40 cores at 2.5\,GHz)}  & \makecell{Intel 2020.4\\Intel MPI 2019.9 } & -O3 -xCORE-AVX512 -std=gnu++17\\
                           \hline
       \makecell{Jean Zay V100\\IDRIS} & \makecell{4$\times$NVidia Tesla V100 \\ +2$\times$Intel Cascade Lake 6248}   & \makecell{nvcc/CUDA 11.2\\OpenMPI 4.1.1} & \makecell{-O3  -expt-extended-lambda\\-arch=sm\_70 -std=c++17} \\
       \hline
       \makecell{AdAstra\\CINES} & \makecell{4$\times$AMD Mi250X \\ +2$\times$AMD EPYC 7713} & \makecell{Cray PE 2.7.19\\Cray Mpich 8.1.21\\Rocm 5.2.0} &\makecell{-O3 -fno-gpu-rdc -x hip \\ --offload-arch=gfx90a -std=gnu++17}\\
       \hline
       
    \end{tabular}

    \label{tab:config}
\end{table*}

\begin{table}[]
    \centering
    \caption{Single node performance and energy efficiency comparison on a variety of architectures. The test problem is the 3D Orszag-Tang vortex test, with a resolution of $32^3$ in each sub-domain on CPUs and $256^3$ in each sub-domain on GPUs. }
    \begin{tabular}{|c|c|c|c|c|c|}
    \hline
       Cluster/ Processor & \makecell{Single node \\performances \\($10^6$ cell/s)} & \makecell{Energy efficiency \\($10^{10}$ cell/kWh)} \\
       \hline
       Irene Rome (CPU)   & 41.3 & 2.7\\
       Jean-Zay CSL (CPU) & 25.0 & 3.0\\
       Jean-Zay V100 (GPU) & 484 & 11.3 \\
       Adastra Mi250 (GPU) & 1240 & 18.2\\
       \hline
       
    \end{tabular}
    
    \label{tab:nodePerf}
\end{table}

 We first present the single node performances in Tab.~\ref{tab:nodePerf} along with an estimation of the energy efficiency. The code typically performs better than 10 Mcell/s on a single Intel 20 cores CPU and 20 Mcell/s on a single AMD 64 cores CPU, which is similar to other CPU codes on the market. When we compare Intel CSL nodes to GPU nodes, we find an acceleration of a factor 19.3 on NVidia V100 and 49.6 on AMD Mi250, demonstrating that performance portability is extremely good.

\begin{figure*}[t!]
   \centering
    \includegraphics[width=1.0\hsize]{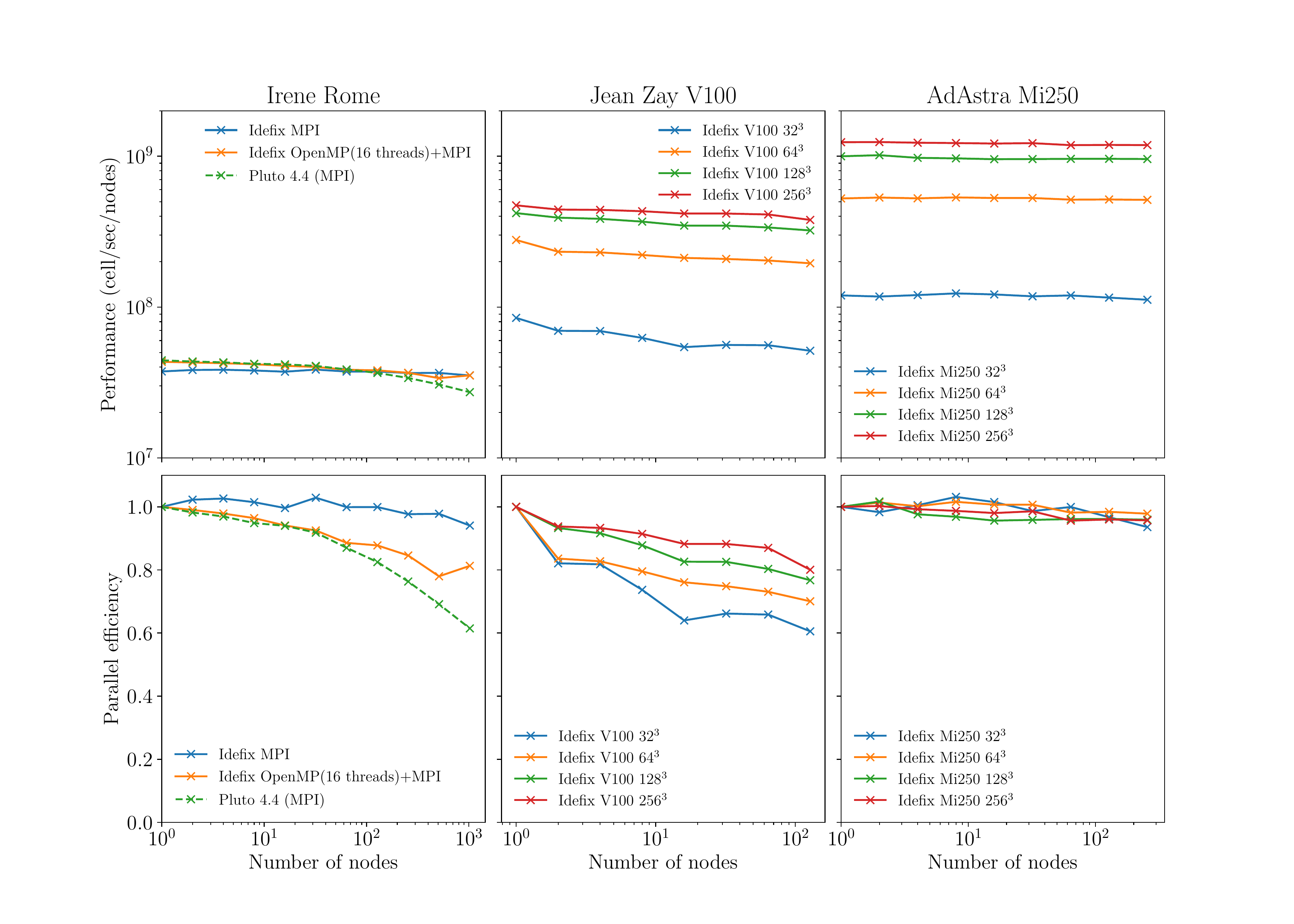}
      \caption{Parallel efficiency (weak scaling) on AMD Rome CPUs (left), NVidia V100 GPUs (middle) and AMD Mi250 GPUs (right). The scaling of \idefix is above 80\% efficiency on up to 131\,072 CPU cores (1024 AMD Rome nodes) and reaches 95\% on 1024 AMD Mi 250 GPUs (256 nodes).}
         \label{fig:scaling}
   \end{figure*}

 Next, we move to the parallelism efficiency of the code, by running the same problem on up to 512 NVidia GPUs (128 nodes), 1024 AMD GPUs (256 nodes) and 131 072 CPU cores (1024 nodes). Measured performances are presented in Fig.~\ref{fig:scaling} for various resolutions. On CPUs, we find that scaling and overall performance is very close to \pluto's. This is not surprising as the algorithms are similar. We note however that \idefix performs slightly better beyond 10 000 cores, which is probably explained by the packing/unpacking strategy when exchanging boundary elements used in \idefix (see section \ref{sec:mpi}). On GPUs, \idefix exhibits an excellent scaling with above 80\% of parallelisation efficiency with subdomains of $256^3$ on Nvidia V100 (Jean Zay), and remarkably above 95\% on AMD Mi250 (Adastra), probably thanks to the AMD Slingshot interconnect used in the latter. We note that performances drop substantially for sub-domains smaller than $64^3$. This is because GPUs use a latency hiding strategy, where memory accesses (which are slow) to some portion of an array are performed simultaneously with computations. This strategy is efficient if there are enough computations to be performed while fetching data in memory, which is not the case below $64^3$ resolution on each GPU. Overall, \idefix achieves a peak performance of about $3\times 10^{11}$ cell/s on 1024 GPUs on Adastra, with sufficiently large sub-domains.
 
 In order to estimate the energy consumption of the node, we have used the Thermal Design Power (TDP) published by the manufacturer of each CPU/GPU. For GPU nodes, we count both the GPU TDP and that of the host CPUs. We find that energy efficiency on CPUs is about $3\times 10^{10}$ cell/kWh while it gets as high as $1.8\times 10^{11}$ cell/kWh on Mi250 GPUs. This demonstrates that for the same problem, \idefix can be up to 6 times more energy efficient on GPUs than on CPUs. We note that our approach is only a proxy for the real energy consumption of the node, as it does not include storage, network and cooling in the energy budget. Moreover, this gain is observed for $256^3$ sub-domain size, and it drops severely as one reduces the domain size: for $32^3$ domain size on GPUs, we get a similar efficiency between CPUs and GPUs. Hence, from an energy sobriety standpoint, the usage of GPUs makes sense only for subdomains larger than $64^3$.
  
\section{Discussions and Conclusions}
In this paper, we have presented \idefix, a new performance-portable code for compressible magnetised fluid dynamics. The code follows a standard finite-volume high-order Godunov approach, where inter-cell fluxes are computed by dedicated Riemann solvers. For MHD problems, \idefix uses constraint transport to evolve the magnetic field  components on cell faces, satisfying the solenoidal condition at machine precision. The code can handle cartesian, polar, cylindrical and spherical coordinate systems with variable grid spacing in every direction. It also features several time integrators including explicit second and third-order Runge-Kutta scheme and Runge-Kutta-Legendre super time-stepping for parabolic terms (viscosity, thermal, ambipolar and ohmic diffusion). A conservative orbital-advection scheme (Fargo-type) is also implemented to accelerate the computation of rotation-dominated flows. The code is parallelised using domain decomposition and the MPI library. Finally, several additional physical modules are currently in development for \idefix, including a multi-geometry self-gravity solver and a particle-in-cell module, which will be discussed in dedicated papers.

All features in \idefix are implemented in a performance-portable way, meaning that they are available for every target architecture supported by \kokkos (CPU or GPU). In contrast to other codes where only a fraction of an algorithm is running on a GPU, \idefix's philosophy is that all data structures and loops live in device space, while the host system is only responsible for input and output operations. Benchmarks show that \idefix performs similarly to \pluto  and other similar Godunov codes on CPU architectures, with an efficiency above 80\% on up to 131 072 CPU cores. On GPU, the speed-up is significant (an AMD Mi250 node being 50$\times$ faster than an Intel Cascade Lake CPU node), and the code is also significantly more energy efficient, up to a factor of about 6, comparing Intel Cascade Lake CPUs to AMD Mi 250 GPUs. However, these performances are only achievable for sufficiently large sub-domain (typically more than $64^3$), which is an intrinsic limitation of GPU architectures. Finally, \idefix reaches peak performance of $3\times 10^{11}$ cell/s in MHD on AdAstra at Cines (1024 GPUs), with a parallelisation efficiency above 95\%.

\idefix is not the only code that proposes performance portability. \textsc{K-Athena} \citep{Grete.Glines.ea19} is similar in essence to \idefix, but is limited to cartesian and ideal-MHD problems. For these problems, the performances of \idefix and \textsc{K-Athena} are quantitatively similar, however, \idefix allows the user to address more complex physics. \textsc{Parthenon} \citep{Grete.Dolence.ea22} is a \kokkos-based AMR framework on which several Godunov codes are being developed, including \textsc{PKAthena}, which would be similar to \idefix. \textsc{Parthenon}'s performances have only been reported for hydro problems, and are similar to \idefix. We note however that \textsc{Parthenon} provides an AMR framework that is not present in \idefix. It is not clear at this stage whether \textsc{Parthenon} will support non-cartesian geometry and all of the features present in \idefix in the future, but \idefix exhibits some clear advantages at the moment.

 The current development of \idefix is tracked on a private git repository, from which a public version\footnote{https://github.com/idefix-code/idefix} is forked. The public version is progressively enriched by new modules once they are published and is distributed under a CECILL license (French Open Source License). \idefix development is done in a collaborative way with modern versioning tools (git). The code master branches (public and private) are validated daily for CPU and GPU targets on a large number of tests (including the ones presented in this paper). Merge requests can be accepted only if the test pipeline fully succeeds, which helps preserve code quality and avoids regression. Finally, the online documentation is generated nightly from the code sources and is available from the code repositories (public and private). While \idefix was initially developed for our own needs in the ERC project "MHDiscs", we believe that the code is now mature enough to be used more broadly and to benefit from external inputs.

\begin{acknowledgements}
We wish to thank our referee, James Stone, for his valuable comments and suggestions that significantly improved the manuscript. We also thank Philipp Grete, Jeffrey Kelling, R\'emi Lacroix, Thomas Padioleau and Simplice Donfack who mentored and helped us improve \idefix during the 2021 hackathon organised by IDRIS and NVidia. Finally, we acknowledge the inspiring work of Pierre Kestener (CEA) on Kokkos-based Godunov methods. This project has received funding from the European Research Council (ERC) under the European Union’s Horizon 2020 research and innovation program (Grant Agreement No. 815559 (MHDiscs)). This work was supported by the "Programme National de Physique Stellaire" (PNPS), "Programme National Soleil-Terre" (PNST), "Programme National de Hautes Energies" (PNHE) and "Programme National de Planétologie" (PNP)  of CNRS/INSU co-funded by CEA and CNES. This work was granted access to the HPC resources of IDRIS and TGCC under the allocation 2022-A0120402231, and a grand challenge allocation at CINES made by GENCI. Some of the computations presented in this paper were performed using the GRICAD infrastructure (https://gricad.univ-grenoble-alpes.fr), which is supported by Grenoble research communities.
      Data analysis and visualisation in the paper were conducted using the scientific Python ecosystem, including numpy \citep{Harris.ea20}, scipy \citep{Virtanen.ea20} and matplotlib \citep{Hunter.07}.
\end{acknowledgements}

\bibliographystyle{aa} 

\bibliography{zutilo.bib}

\end{document}